\documentclass[12pt,preprint,a4paper]{aastex}
\usepackage{graphics,epsf}
\usepackage{amsmath}                
\usepackage{amsfonts}               
\usepackage{amssymb}                
\usepackage{epsfig}                 
\usepackage{subfigure}
\usepackage{float}

\def \kms{$\rm{km}~\rm{s}^{-1}$}

\def \cm{~\rm{cm}}
\def \s{~\rm{s}}
\def \km{~\rm{km}}

\def \K{~\rm{K}}
\def \g{~\rm{g}}

\def \erg{~\rm{erg}}
\def \yrs{~\rm{yrs}}
\def \yr{~\rm{yr}}
\def \pc{~\rm{pc}}

\begin{document}

\title{TYPE Ia SUPERNOVAE INSIDE PLANETARY NEBULAE: SHAPING BY JETS}

\author{Danny Tsebrenko\altaffilmark{1} and  Noam Soker\altaffilmark{1}}

\altaffiltext{1}{Department of Physics, Technion -- Israel Institute of Technology, Haifa 32000, Israel;
ddtt@tx.technion.ac.il; soker@physics.technion.ac.il.}
\begin{abstract}
Using 3D numerical hydrodynamical simulations we show that jets launched prior to type Ia supernova (SN Ia)
explosion in the core-degenerate (CD) scenario can account for the appearance of two opposite lobes ('Ears') along
the symmetry axis of the SN remnant (SNR).
In the double-degenerate (DD) and CD scenarios the merger of the two degenerate compact objects
is very likely to lead to the formation of an accretion disk,
that might launch two opposite jets.
In the CD scenario these jets interact with the envelope ejected during the preceding common envelope phase.
If explosion occurs shortly after the merger process, the exploding gas and the jets will collide with the ejected nebula,
leading to SNR with axisymmetric components including 'Ears'.
We also explore the possibility that the jets are launched by the companion white dwarf prior to its merger with the core.
This last process is similar to the one where jets are launched in some pre-planetary nebulae.
The SNR 'Ears' in this case are formed by a spherical SN Ia explosion inside an elliptical planetary nebula-like object.
We compare our numerical results with two SNRs - Kepler and G299.2-2.9.
\end{abstract}
\section{INTRODUCTION}
\label{sec:intro}
Type Ia Supernovae (SNe Ia) are thermonuclear detonations of carbon-oxygen white dwarfs (WDs;{ }\citealt{Hoyle1960}).
The exact mechanism leading to the explosion is debated.
Four viable scenarios for SNe Ia are currently considered:
(a) the single degenerate (SD) scenario (e.g., \citealt{Whelan1973, Nomoto1982, Han2004}),
in which a WD accretes mass from a non-degenerate stellar companion;
(b) the double degenerate (DD) scenario (e.g., \citealt{Webbink1984, Iben1984}), in which a merger of two WDs takes place.
Recent papers discuss violent merger and collision (e.g.,
\citealt{Pakmor2013, Kushniretal2013}) as an ignition channel of the DD scenario;
(c) the core-degenerate (CD) scenario (e.g., \citealt{Livio2003, Kashi2011, Soker2011, Ilkov2012, Ilkov2013, Soker2013}),
in which a WD merges with a hot core of a massive asymptotic giant branch (AGB) star;
(d) the 'double-detonation' mechanism (e.g., \citealt{Woosley1994, Livne1995}),
in which a sub-Chandrasekhar mass WD accumulates on its surface a layer of helium-rich material,
which can detonate and lead to a second detonation near the center of the CO WD (e.g., \citealt{Shenetal2013} for a recent paper).

In the violent merger model \citep{Pakmor2012} it is possible that in the merger process
of the two WDs the helium is ignited first.
In this mechanism both the DD and the double detonation operate \citep{Pakmor2013}.
The double detonation might operate in the CD scenario as well, with or without a violent merger.
In both the SD and the CD scenarios,
a circumstellar shell can be formed by ejecting the companion (to the WD) stellar envelope close to the explosion time.
This ejection can be accompanied by the formation of jets launched from an accretion disk.
In the SD scenario the disk is formed during the mass accretion process onto the WD,
while in the CD scenario the disk is formed either around the core from the destructed WD material,
or around the WD from the destructed core.
The formation of a disk as a result of merger of two WDs was discussed before, e.g.,  \cite{Raskin2013}
 and \cite{Ji2013} for recent papers.
Another possibility in the CD scenario is that the WD accretes mass from the giant stellar envelope
and launches two opposite jets before merging with the giant's core.
Such circumstellar matter (CSM) shells are similar to some planetary nebulae (PNs), as the central WD ionizes the CSM.
When the SN is finally ignited, the WD mass in the SD scenario and the merger product mass in the CD scenario
is expelled by the explosion at high velocities of up to $V_{\rm SNm}=20,000 \km \s^{-1}$.
If the explosion occurs within $\sim 10^5 \yrs$ from the CSM ejection episode,
we have a SN Ia exploding inside a PN or a PN-like object.
The exploding gas catches up with the dense (relative to the interstellar medium; ISM) CSM and interacts with it.

In this paper we study the interaction of SN Ia ejecta with a PN (or PN-like) shell.
Observations show that some of the supernova remnants (SNRs) exhibit non-spherical morphologies,
including axi-symmetrical morphologies resembling the structures of some PNe.
An example is the Kepler SNR that possesses two opposite protrusions
which are termed `Ears' (after the nomenclature used in \citealt{Cassam2004}).
Another example is the G299.2-2.9 SNR.
For the general discussion (not limited to Kepler or G299.2-2.9 SNRs), we will refer to these protrusions as 'ears'.

\begin{figure}[h!]
\begin{center}
\subfigure{\label{subfigure:kepler}\includegraphics*[scale=0.3,clip=true,trim=0 0 0 0]{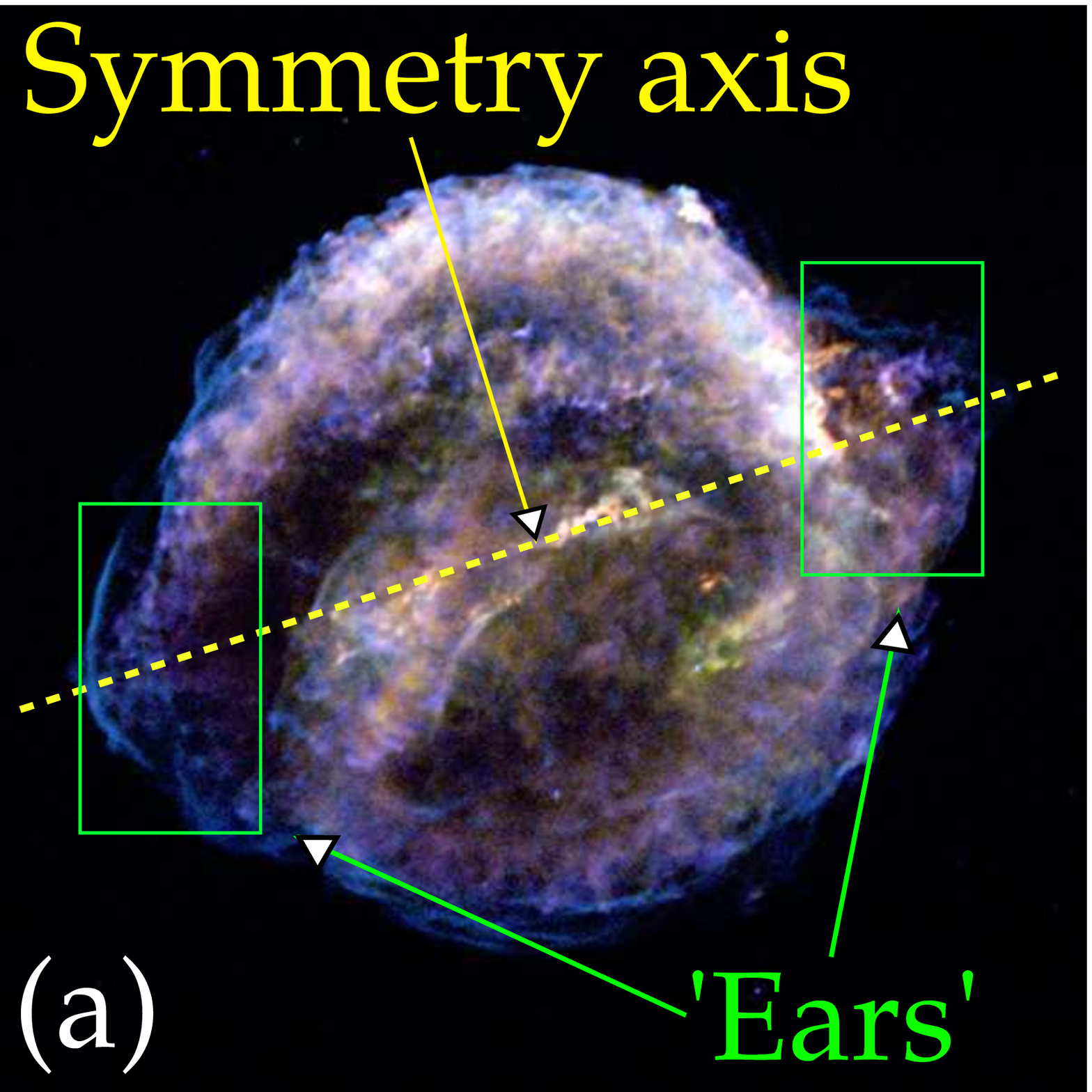}}
\hspace{1.0 cm}
\subfigure{\label{subfigure:299}\includegraphics*[scale=0.3,clip=false,trim=0 0 0 0]{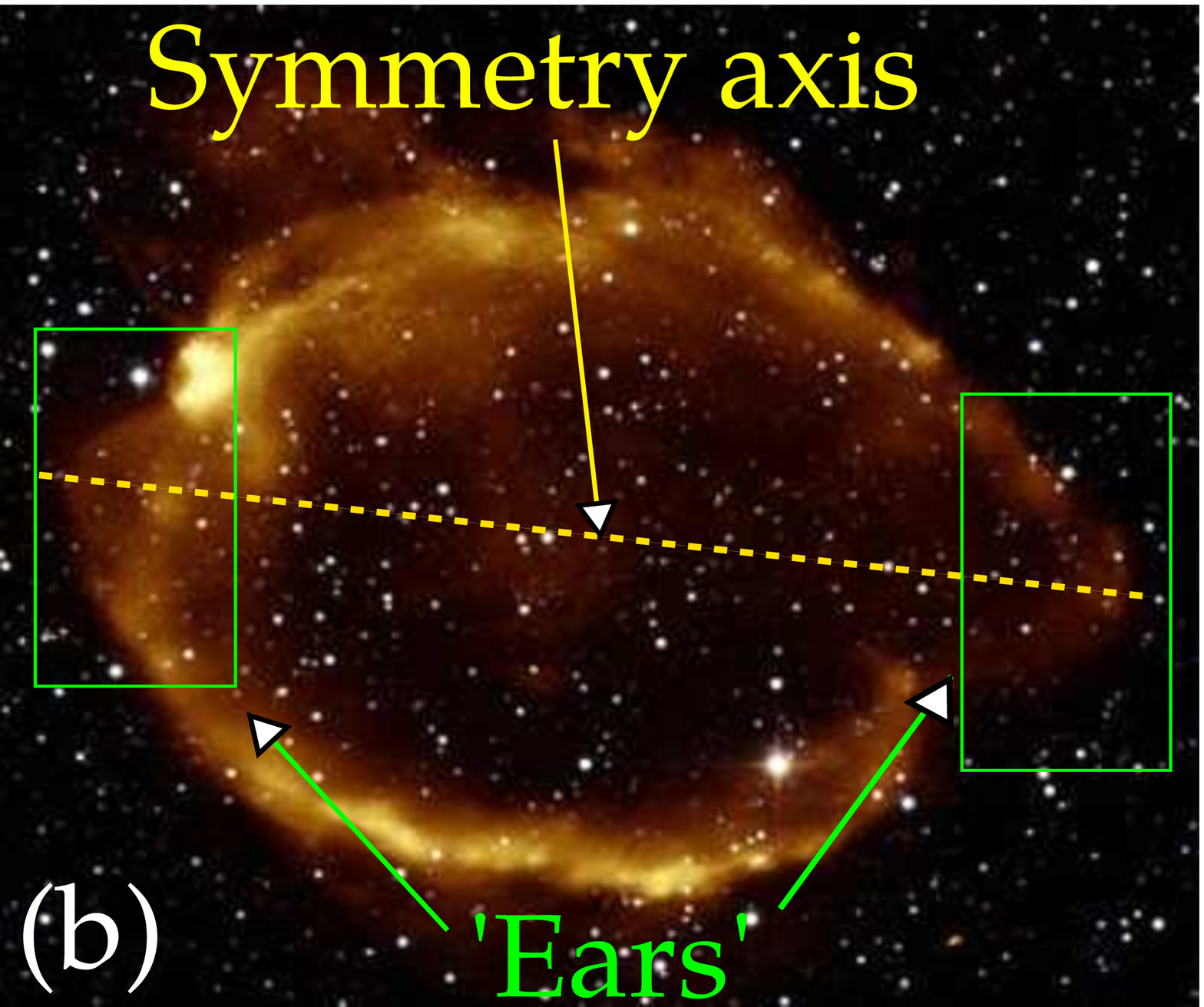}}
\caption{
(a) The Kepler SNR. Merged image between 0.3 and 8 keV taken from \cite{Reynolds2007}.
Note that our morphological interpretation of the Kepler SNR is different by $90^\circ$ than that of \cite{Burkey2013},
who take the equatorial plane to be where we take the symmetry axis on the figure (dashed line).
(b) G299-2.9 SNR. Merged image between 0.3 and 3 keV taken from \cite{Park2007}.
Marked in both images are the observed blob-like features ('Ears'),
which we attribute to jets blown either a long time before the explosion or immediately before the explosion.}
\label{fig:observations}
\end{center}
\end{figure}

The motivation for this work comes from the expectation that jets will be formed in some SN Ia scenarios,
and may suggest an explanation to the observed two 'Ears' of the Kepler SNR and the G299.2-2.9 SNR.
The 'Ears' are marked on an the X-ray images of these two SNRs shown in Fig. \ref{fig:observations}.
We perform 3D hydrodynamical simulations of such SN explosions
and study the interaction of the expelled material with the CSM.
The numerical code and initial setting and conditions are described in Section \ref{sec:numerical},
while in Section \ref{sec:results} we demonstrate that the double 'ears' and similar morphological features
can be explained by jets that are launched during the pre-explosion phase.
Our discussion and comparison with the SNRs Kepler and G299.2-2.9 are in Section \ref{sec:comparison},
and a short summary is given in Section \ref{sec:summary}.
\section{NUMERICAL SETUP}
\label{sec:numerical}
The simulations are performed using the high-resolution multidimensional hydrodynamics code {\sc{flash 4.0}} \citep{Fryxell2000}.
We employ a full 3D uniform grid (all cells have the same size) with Cartesian $(x,y,z)$ geometry.
In all runs but one each axis has 512 cells and a total length of $\Delta=3.2 \pc$.
In one run each axis still has 512 cells but $\Delta=0.65 \pc$.
We use radiative cooling for solar metallicity values, according to \cite{Sutherland1993},
although radiative cooling has a minor role in our simulations.
The ejected material is metal-rich (see a recent paper by \citealt{Park2013})
and the cooling of the shocked ejecta is expected to be more
efficient than what the cooling function mentioned above gives.
However, this difference in not significant for the duration of our simulations and
does not play a significant role in the shaping of the 'Ears'.

There are two scenarios for forming various shapes of the CSM during the stage prior to SN Ia explosion.
In the SD scenario the WD grows in mass through accretion from a non-degenerate stellar companion.
If the companion is an evolved AGB star it might blow large parts of its envelope, as progenitors of PNe do.
The accretion onto the WD takes place via an accretion disk, which enables jets to be formed.
The jets can blow two opposite small lobes ('ears') in the nebula as observed in some PNe.
Examples of elliptical PNe with 'ears' are NGC~6563 \citep{Schwarz1992}, NGC~7139, IC~418, and NGC 3242
(\citealt{Balick1987}; their images can be seen in the
Planetary Nebula Image Catalogue\footnote{http://www.astro.washington.edu/users/balick/PNIC/}).
We note that in some cases the two 'ears' are not exactly symmetric to each other.
As well, in some PNe there are indications that the 'ears' are formed by jets.

In the CD scenario \citep{Soker2013} the common envelope (CE) phase is
terminated by a merger of a WD companion with the hot core of a massive AGB star.
However, in the cases where the merger takes place when the WD is denser than the core,
the core will be disrupted and accreted onto the cooler WD.
In such cases the explosion might occur shortly, months to tens of years, after the merger process.

Whereas in the SD scenario the shaping of 'ears' by jets is only of the CSM shell,
in the CD scenario the shaping can occur both in the CSM and/or during the core-WD merger.
In the rare cases when the SN explosion occurs within $\simeq 10^5 \yrs$ from the formation of the CSM,
jets can be formed prior to the explosion and, together with the high-velocity SN ejecta, collide with the CSM
and lead to formation of specific features in the SNR.
This type of interaction might form SN Ia remnants that in general possess a large scale spherical symmetry,
but have small, notably axisymmteric, departures from sphericity.
We examine whether the formation of the 'Ears' in the SNRs of Kepler and G299.2-2.9
can be explained by such process.

We note that protrusions can also be formed by 'bullets' ploughing through the shell,
as suggested by recent simulations of Tycho's SNR and SN 1006 \citep{Warren2013}.
\cite{Warren2013} show that knots of SN ejecta can overtake the forward shock and form low density bubbles around the rim of the SNR.
However, these features appear only for one specific simulated case having an adiabatic index of $\gamma = 6/5$,
and are absent entirely from other simulations.
We aim at two opposite protrusions in an axisymmetrical structure, rather than at randomly distributed protrusions.

Based on the above discussion we simulate the two scenarios for the formation of protrusions.
One scenario sets the asymmetry in the PN shell long before the explosion,
while the other assumes the launching of two opposite jets very shortly before the explosion.
Based on these, we conduct simulations with two types of initial setups of explosion inside a CSM shell.
(a) {\it CSM-lobes model:} a spherical shell with two hollow small lobes, mimicking the structure of such two opposite small lobes observed in some elliptical PNs.
The initial distribution of the SN ejecta is spherically symmetric.
(b) {\it Pre-explosion jets (PEJ) model:} a completely spherical CSM shell, with two jets added inside the otherwise spherical SN ejecta.
This can occur only in the CD scenario during the core-WD merger process \citep{Soker2013}.

The SN ejecta density is modeled by an exponential density profile 
\citep{Dwarkadas1998}
\begin{equation}
\rho_{\rm{SN}} = A \,{\rm{exp}}{} (-v/v_{\rm{ejecta}})t^{-3},
\end{equation}
where $v_{\rm{ejecta}}$ is a constant which depends on the mass and kinetic energy of the ejecta,
\begin{equation}
v_{\rm{ejecta}} = 2.44 \times 10^8 E_{51}^{1/2} \left(\frac{M_{\rm{SN}}}{M_{\rm{Ch}}}\right)^{-1/2} \cm \s^{-1},
\end{equation}
$M_{Ch} = 1.4 M_\odot$,
$E_{51}$ is the explosion energy in units of $10^{51} \erg$, and $A$ is a parameter given by
\begin{equation}
A = 7.67 \times 10^6 \left(\frac{M_{\rm{SN}}}{M_{\rm Ch}}\right)^{5/2} E_{51}^{-3/2}  \g \s^{3} \cm^{-3} .
\end{equation}
The maximum velocity of the SN ejecta is taken to be $v_{\rm SNm}=20,000 \km \s^{-1}$.
Each simulation starts roughly $8 \yrs$ after the explosion took place.
By this time the fastest ejecta reached $0.16 \pc$ from the center of the explosion.
The total energy of the explosion is set to be $E_{\rm SN}=10^{51} \erg$, and the mass ejected in the explosion is $M_{\rm SN}=1.4 M_\odot$.
The CSM initial profile is set to be a constant density \citep{Patnaude2012},
$\rho_{\rm CSM}=3.15 \times 10^{-21} \g \cm^{-3}$, shell within radii $0.24-0.27 \pc$,
so that the total mass of the CSM shell is $M_{\rm CSM} \simeq 1M_\odot$. The mass is based on the estimated CSM mass in the Kepler SNR \citep{Borkowski1992, Borkowski1994, Kinugasa1999, Blair2007}.
The ambient ISM density is taken to be $\rho_{\rm ISM}= 10^{-24} \g \cm^{-3}$ \citep{Vink2008}.

\begin{figure}[ht]
\begin{center}
\subfigure{\label{subfigure:lobesscheme}\includegraphics*[scale=0.3,clip=true,trim=0 0 0 0]{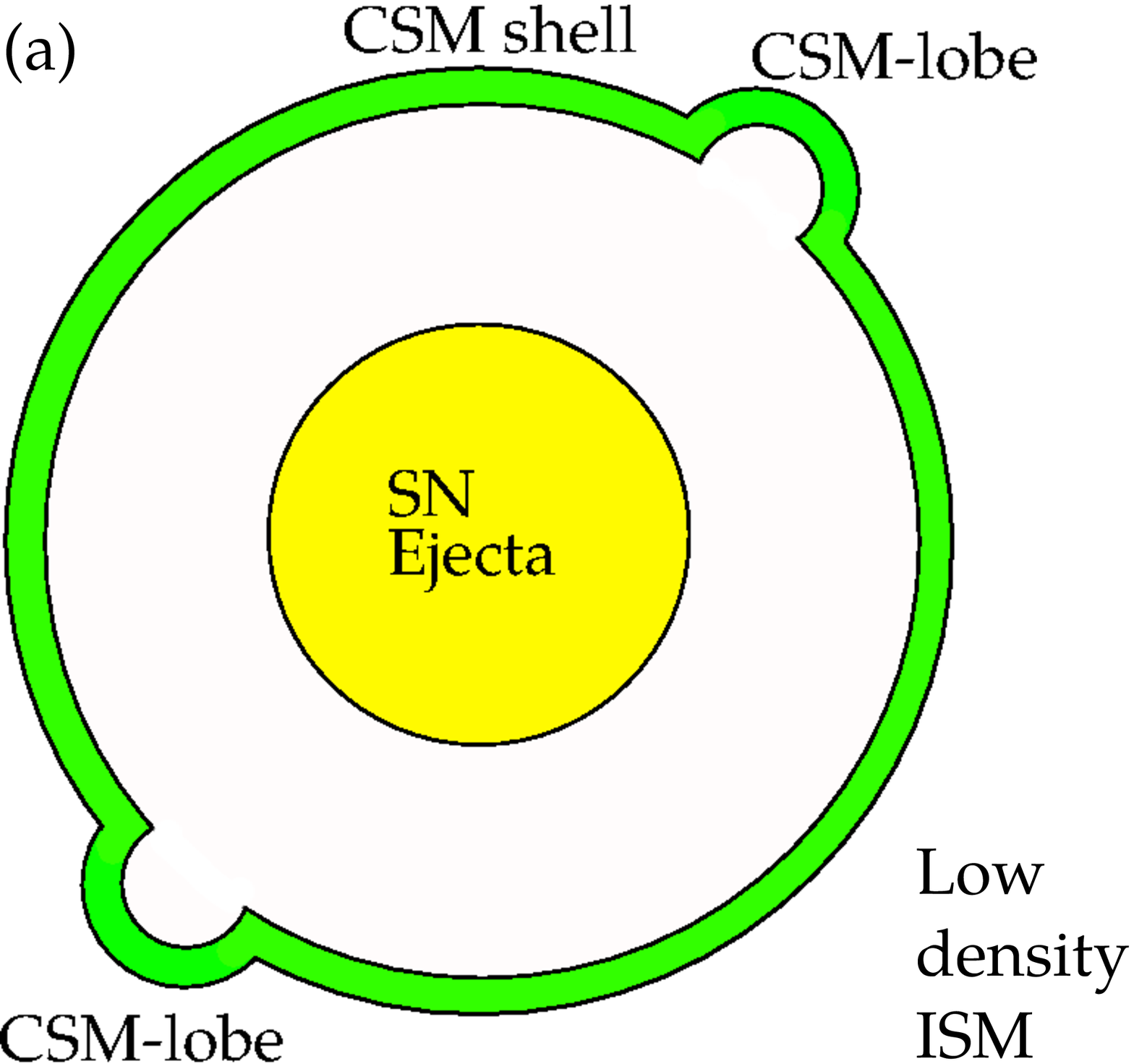}}
\hspace{1.0 cm}
\subfigure{\label{subfigure:jetsscheme}\includegraphics*[scale=0.3,clip=false,trim=0 0 0 0]{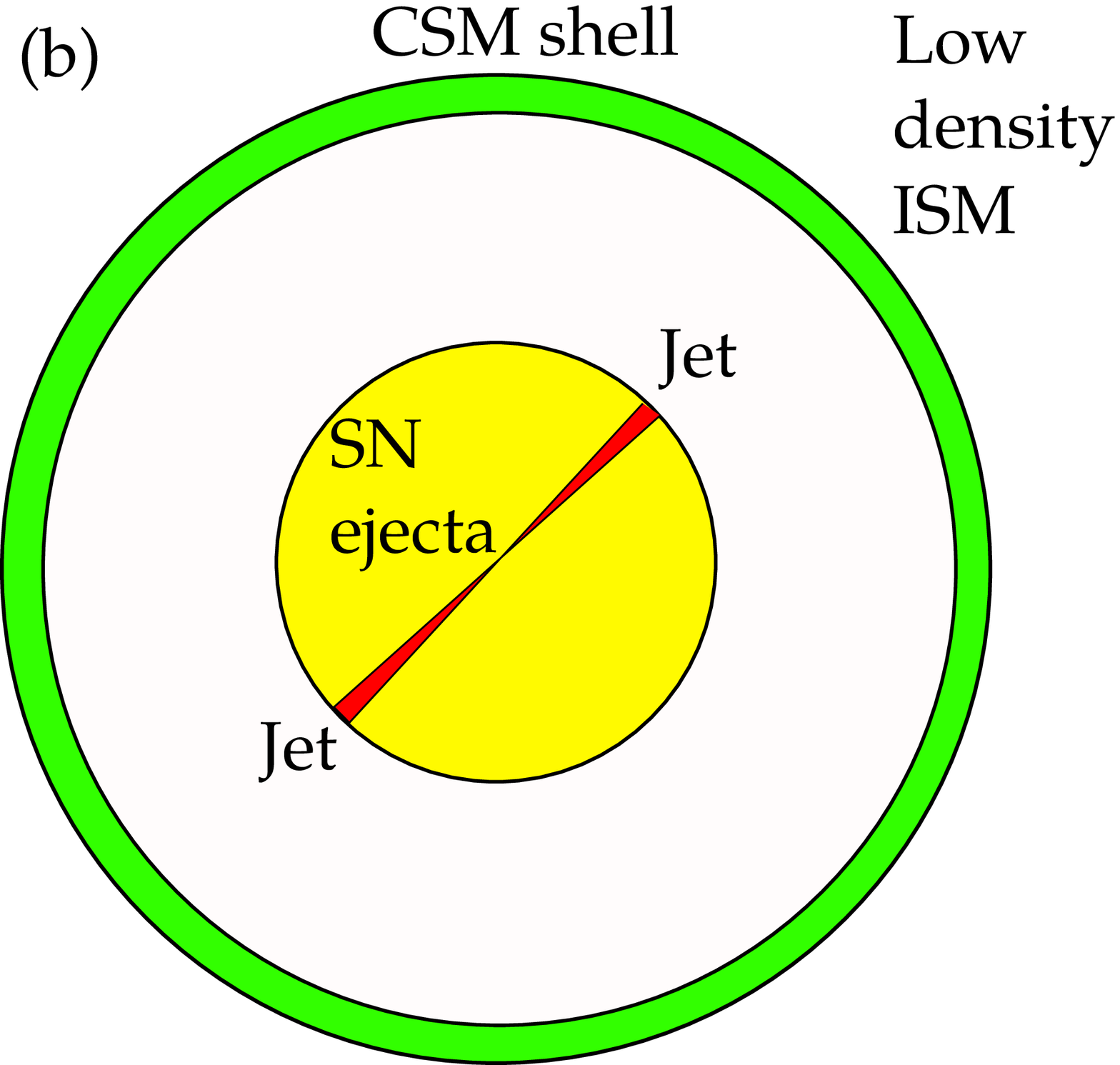}}
\caption{The initial geometry of the circumstellar medium (CSM) and supernova ejecta in the two models. (a) The small CSM-lobes model where two initial lobes are introduced following morphologies of some planetary nebulae. The explosion is spherical. (b) The pre-explosion jets (PEJ) model where jets are launched during the merger process seconds to days prior to the explosion.}
\label{fig:schemes}
\end{center}
\end{figure}

In the small CSM-lobes model, some of the mass of the CSM envelope is 'pushed outward' to form the 'Ears'.
Each small-lobe has an initial half-spherical shell structure within radii $0.065-0.09 \pc$ from the center of the small-lobe.
We assume that the small CSM-lobes are formed by jets with low mass,
and hence take the total mass of the CSM in the small-lobes to be the same as the CSM shell segments they replaced.
As the width of the small lobes is as that of the CSM,
the density in the small lobes is lower than that in the CSM,
and amounts to a value of $\rho_{\rm small-lobes}=1.56 \times 10^{-21} \g \cm^{-3}$.
The initial setup of the small CSM-lobes is shown in Fig. \ref{subfigure:lobesscheme}.

In the PEJ model, jets are added to the explosion itself.
The reason is that in the CD scenario the jets might occur during the core-WD merger \citep{Soker2013},
and the explosion may occur very shortly after the merger, e.g.,
as in the violent merger ignition model of \cite{Pakmor2013}.
The jets are expected to be launched at about the escape velocity from the massive and compact WD,
$v_{\rm jets}> 10,000 \km \s^{-1}$ , not much different from the SN explosion velocity.
To make the model simple, we set the velocity and density radial profiles of the jets to be same as that of the SN ejecta,
but take the density in the jets to be 3 times larger that that of the SN ejecta at the same radius.
The mass of each jet is $ 0.03 M_{\odot}$ and
the half-opening angle of the jets is $\theta = 5^{\rm{o}}$.
The initial setup of the PEJ model is shown in Fig. \ref{subfigure:jetsscheme}.

To have better understanding of the process of interaction of the SN ejecta with the CSM at early times (up to $\simeq 20 \yr$)
we perform one additional simulation that focuses on the region within the initial radius of the CSM shell.
In this setup each axis still has 512 cells but is only $0.65 \pc$ in length,
so that the effective resolution is increased significantly in comparison to the simulation runs described above.

The numerical $x-y$ $(z=0)$ plane is taken to be in the plane containing the central explosion
and the two 'Ears' or jets.
Initially this plane is a symmetry plane, but we simulate the entire space,
so that we assume no symmetry about the $z=0$ axis at $t>0$.
\section{RESULTS}
\label{sec:results}
\begin{figure}[h!]
\begin{center}
\hspace*{-1.0 cm}
\vspace*{-0.5 cm}
\subfigure{\label{subfigure:densears1}\includegraphics[scale=0.42,clip=false,trim=0 0 0 0]{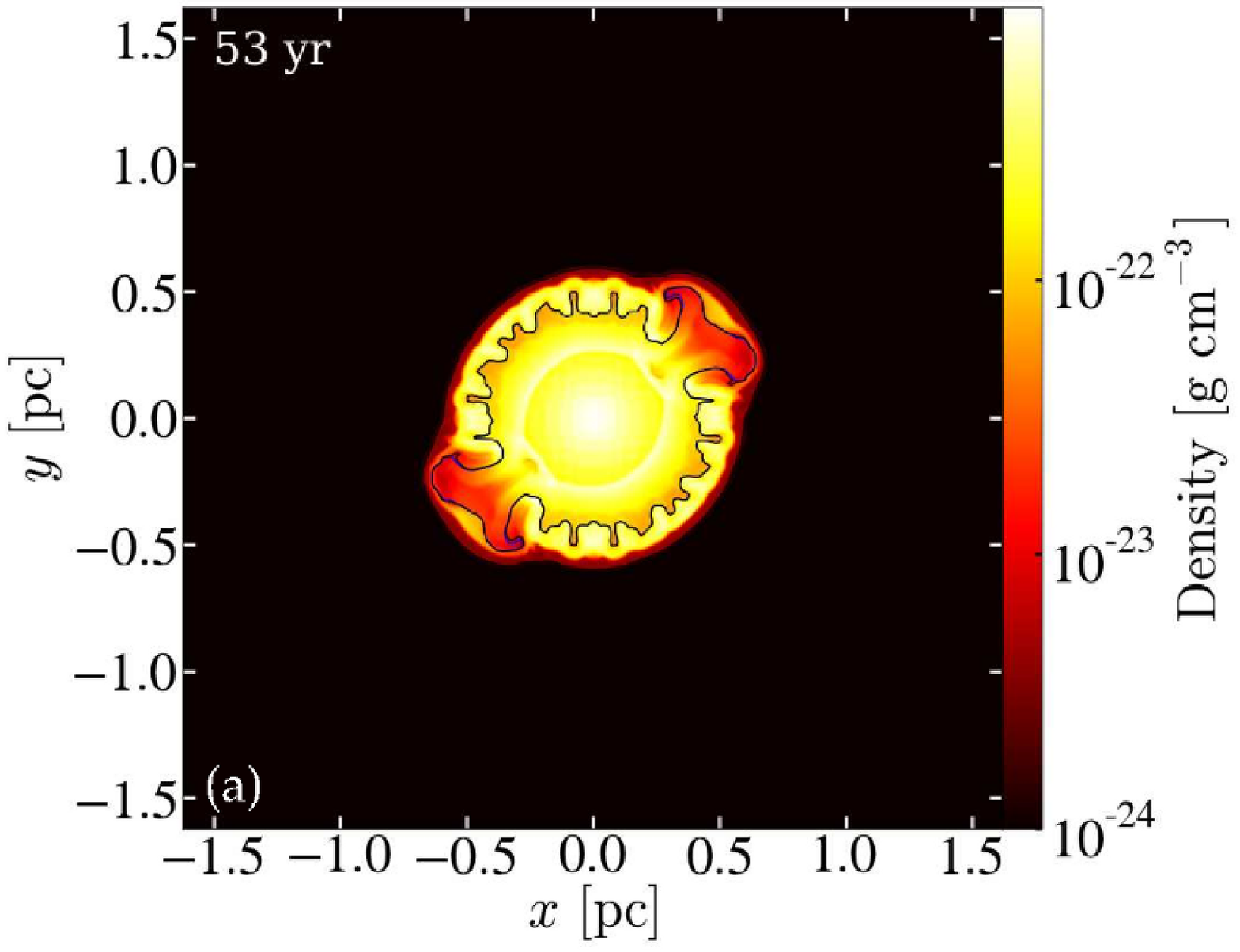}}
\subfigure{\label{subfigure:densears2}\includegraphics[scale=0.42,clip=false,trim=0 0 0 0]{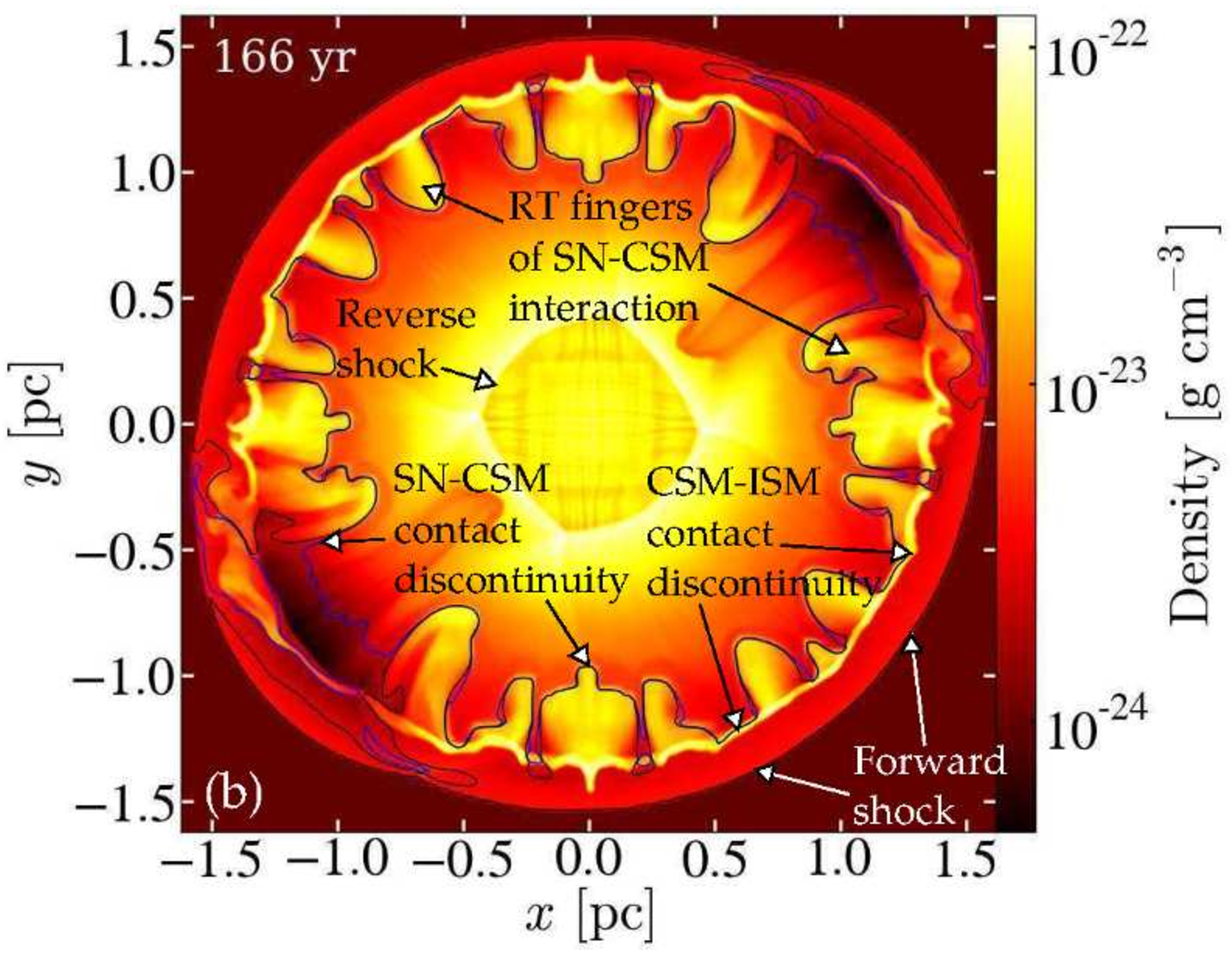}}\\
\hspace*{-1.0 cm}
\vspace*{-0.5 cm}
\subfigure{\label{subfigure:densjets1}\includegraphics[scale=0.42,clip=false,trim=0 0 0 0]{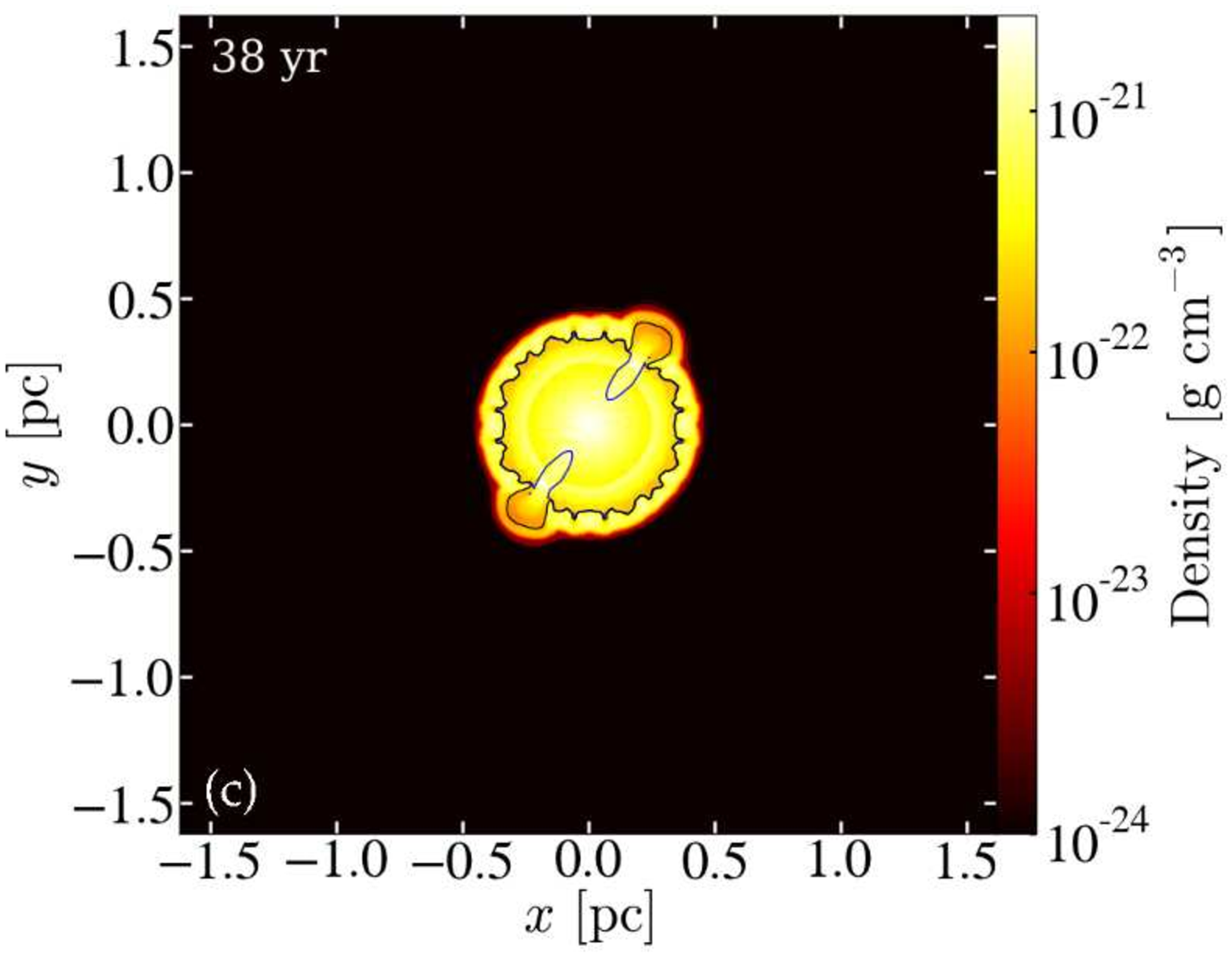}}
\subfigure{\label{subfigure:densjets2}\includegraphics[scale=0.42,clip=false,trim=0 0 0 0]{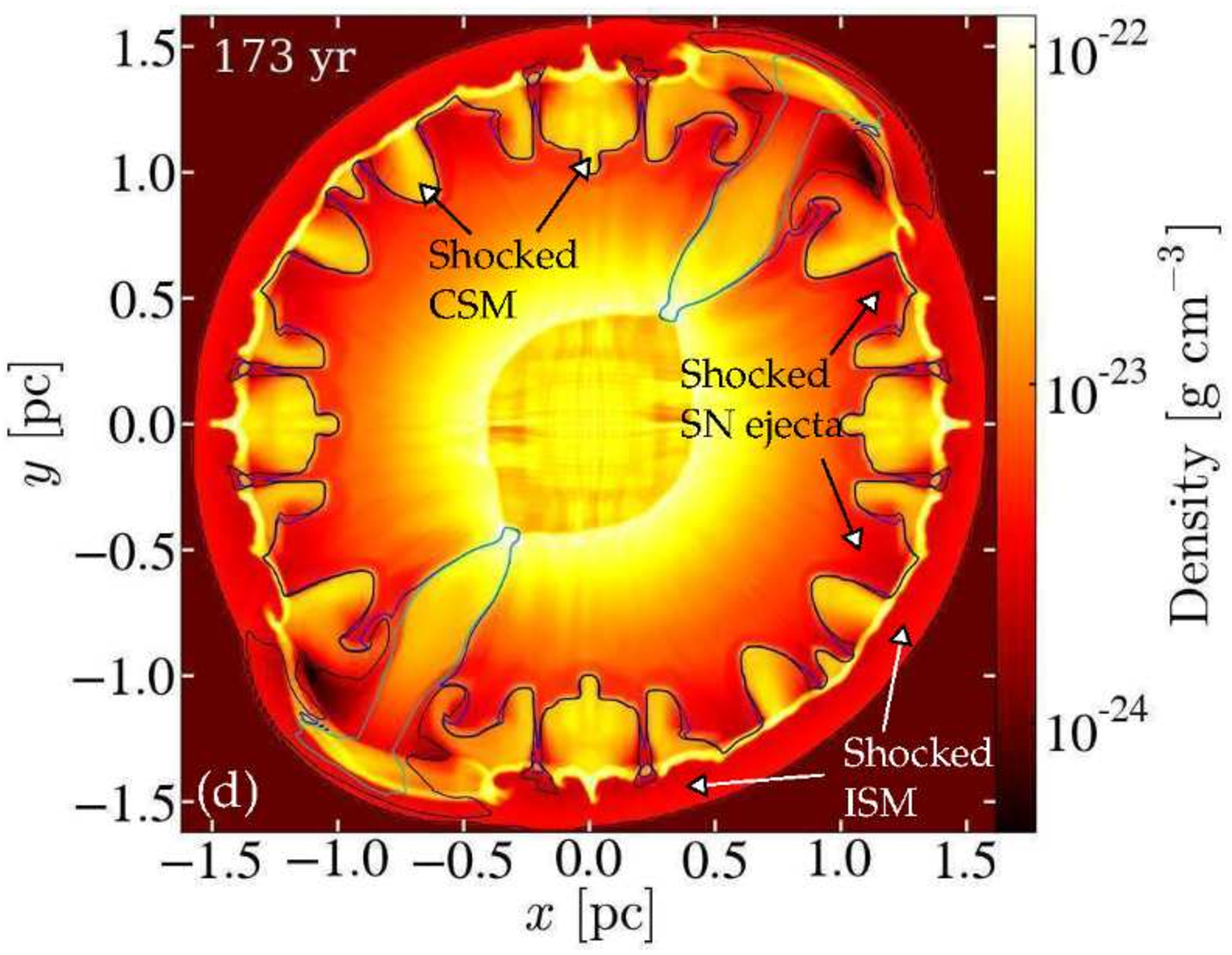}}\\
\caption{(a,b): CSM-lobes model whose initial conditions are depicted in Fig. \ref{subfigure:lobesscheme}.
(c,d): Spherical CSM with pre-explosion jets (PEJ) model whose initial conditions are depicted in Fig. \ref{subfigure:jetsscheme}.
Shown is the density structure (color-coded) at various times as indicated, in the $x-y$ plane.
The contour lines distinguish matter located initially in the jets (cyan; only in the PEJ model),
in the SN ejecta (blue) and in the CSM shell (black).
We emphasize that the ISM might contain significant amount of low-density CSM
from a stellar wind prior to the main CSM shell ejection.
}
\label{fig:densityPlots}
\end{center}
\end{figure}

\begin{figure}[h!]
\begin{center}
\hspace*{-1.0 cm}
\subfigure{\label{subfigure:jetsxy}\includegraphics[scale=0.42,clip=false,trim=0 0 0 0]{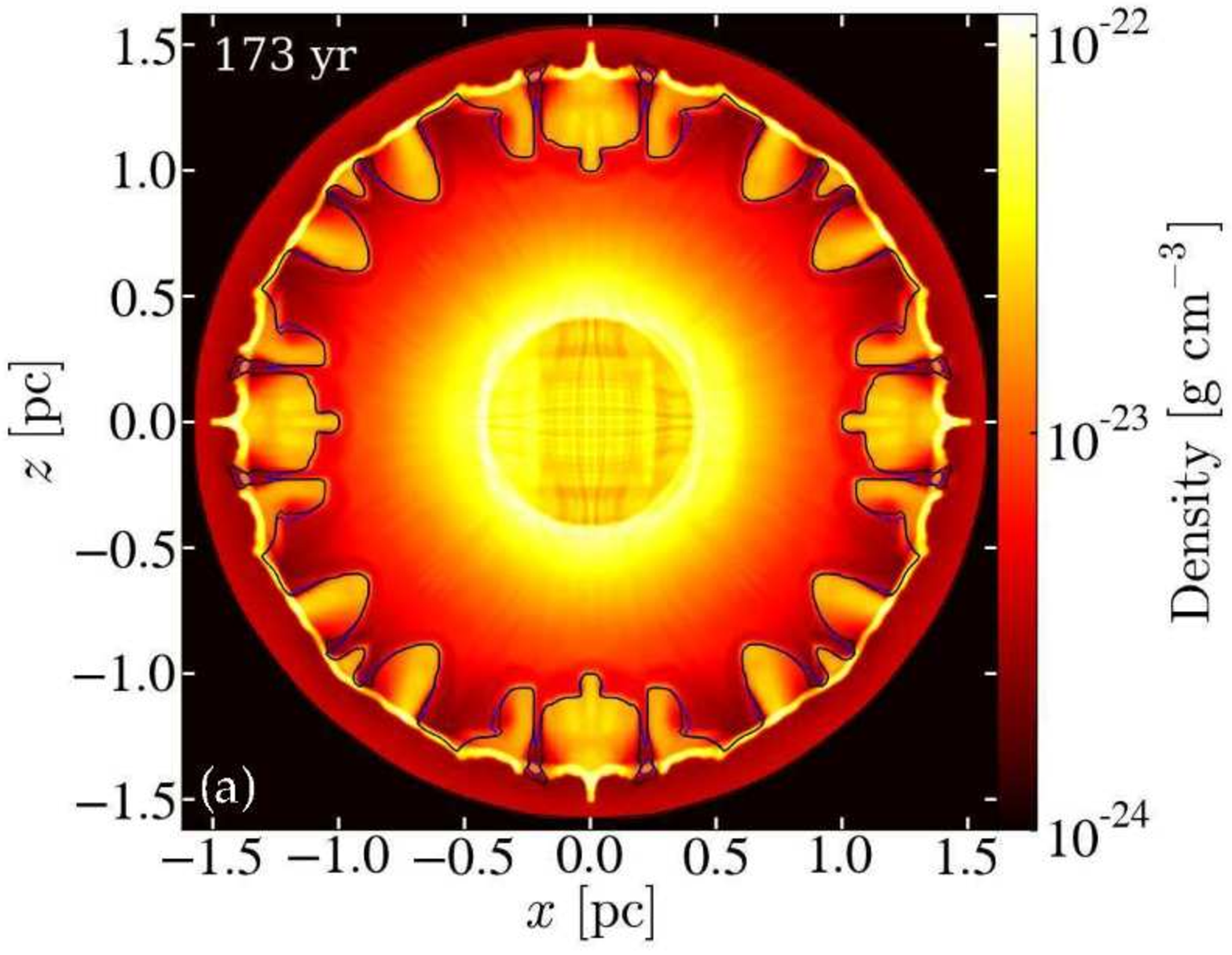}}
\subfigure{\label{subfigure:jetsz=0.5pc}\includegraphics[scale=0.42,clip=false,trim=0 0 0 0]{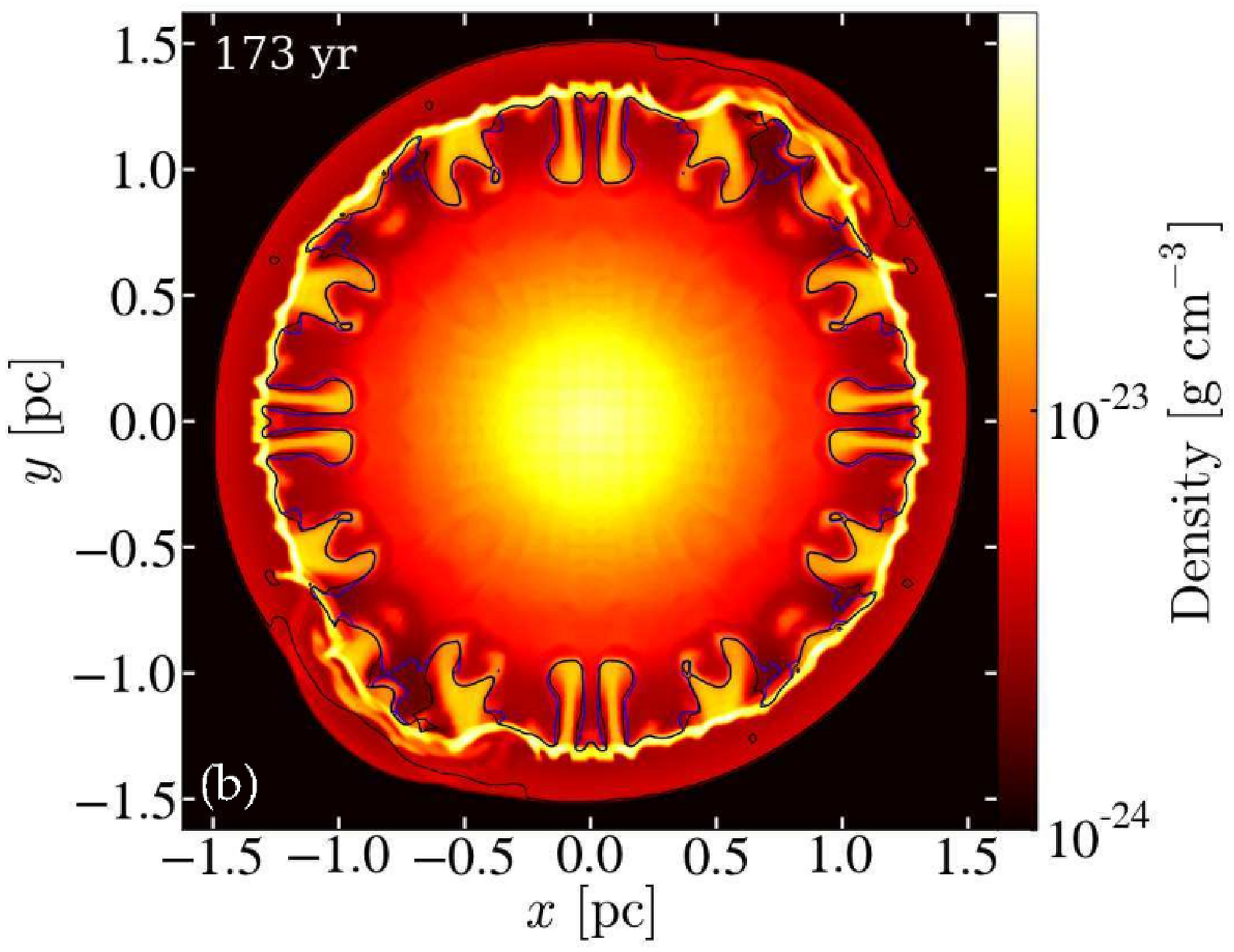}}
\caption{
Same as Fig. \ref{fig:densityPlots}, in two additional planes.
(a) Spherical CSM with PEJ in the $y=0$ $(x-z)$ plane.
(b) Spherical CSM with PEJ in the $z=0.5 \pc$ plane.
}
\label{fig:densityPlots2}
\end{center}
\end{figure}

\begin{figure}[h!]
\begin{center}
\hspace*{-1.0 cm}
\subfigure{\label{subfigure:jetstemperature}\includegraphics[scale=0.39,clip=false,trim=0 0 0 0]{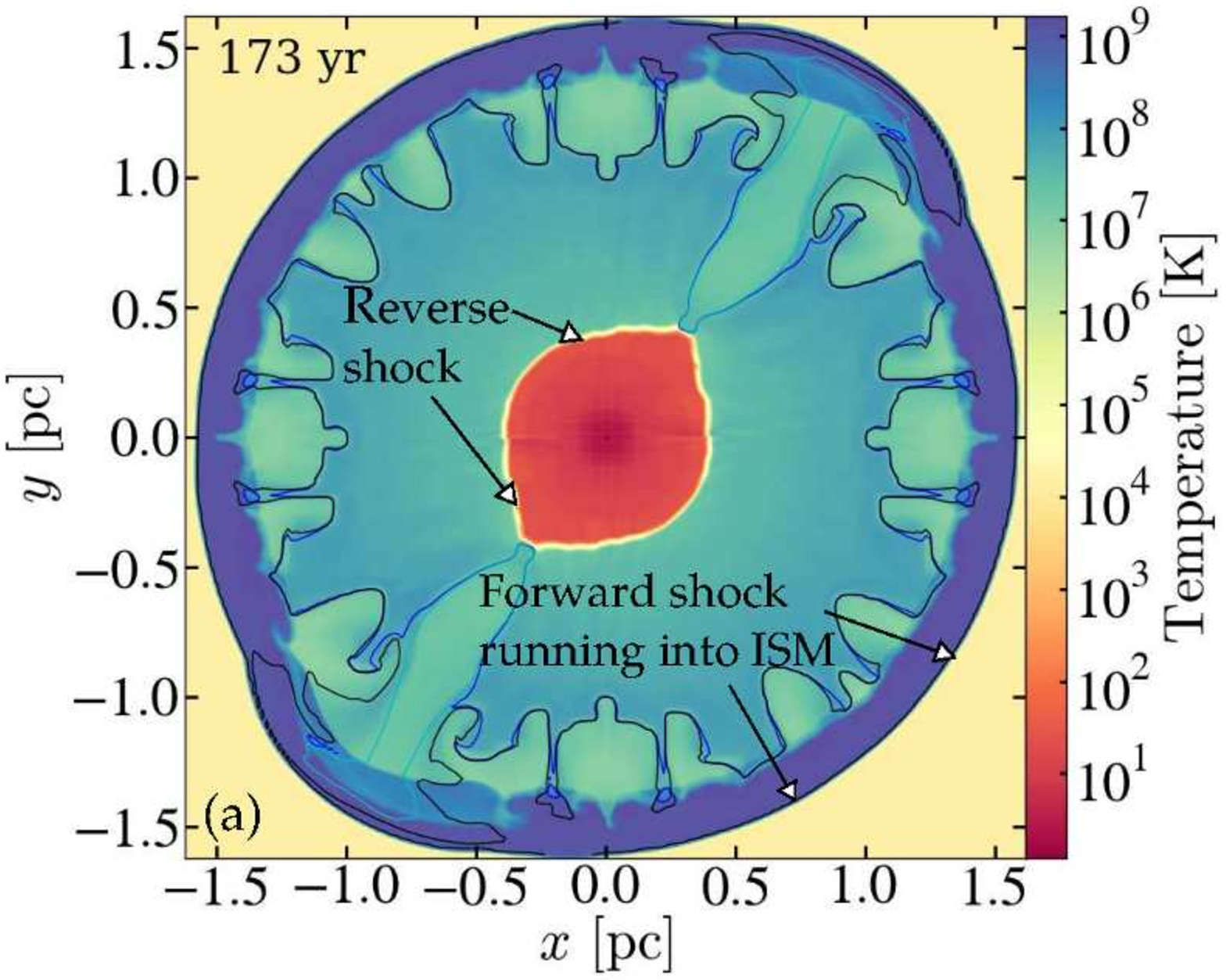}}
\hspace{-0.1 cm}
\subfigure{\label{subfigure:jetspressure}\includegraphics[scale=0.4,clip=false,trim=0 0 0 0]{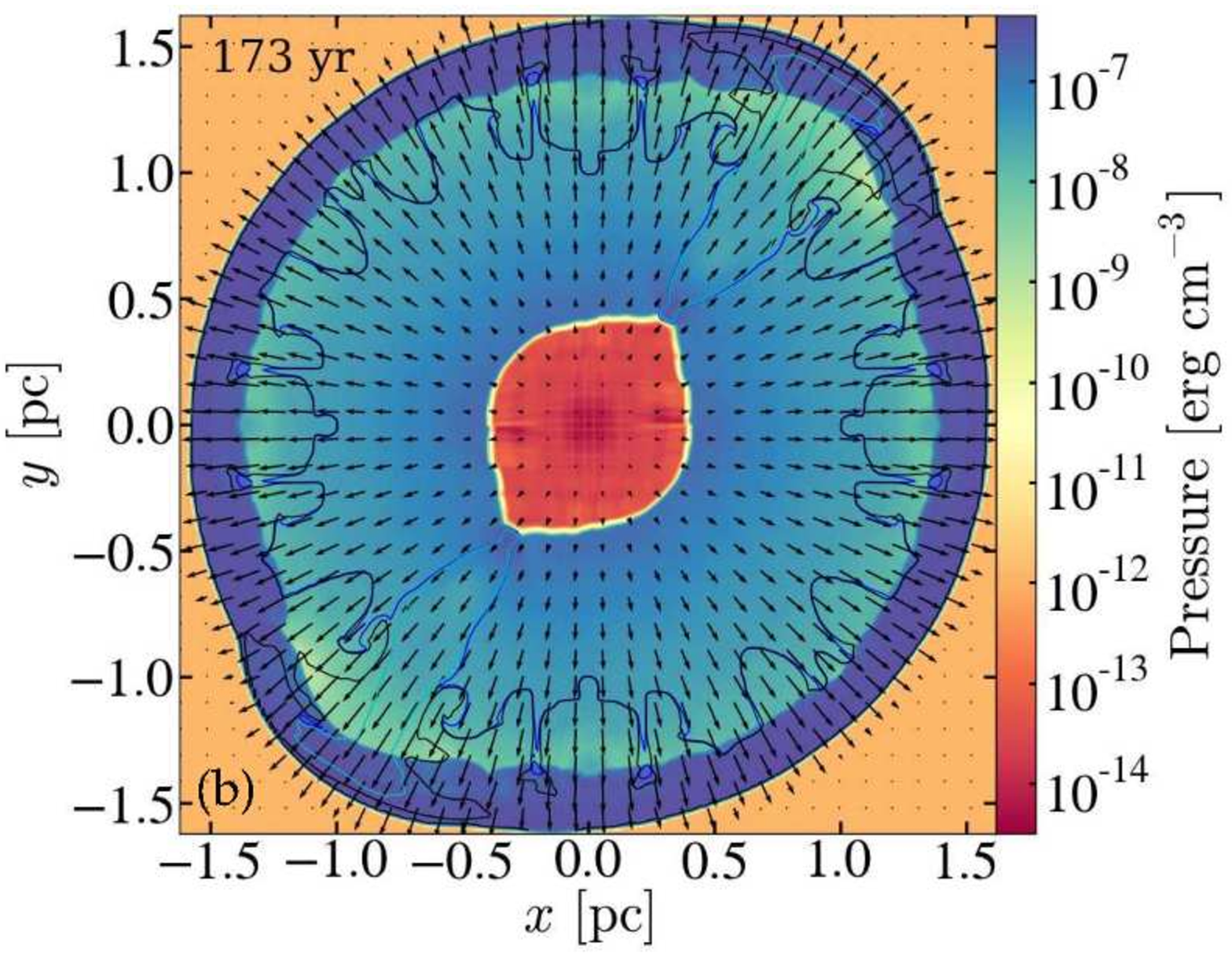}}\\
\hspace*{-0.85 cm}
\subfigure{\label{subfigure:jetsrt1}\includegraphics[scale=0.41,clip=false,trim=0 0 0 0]{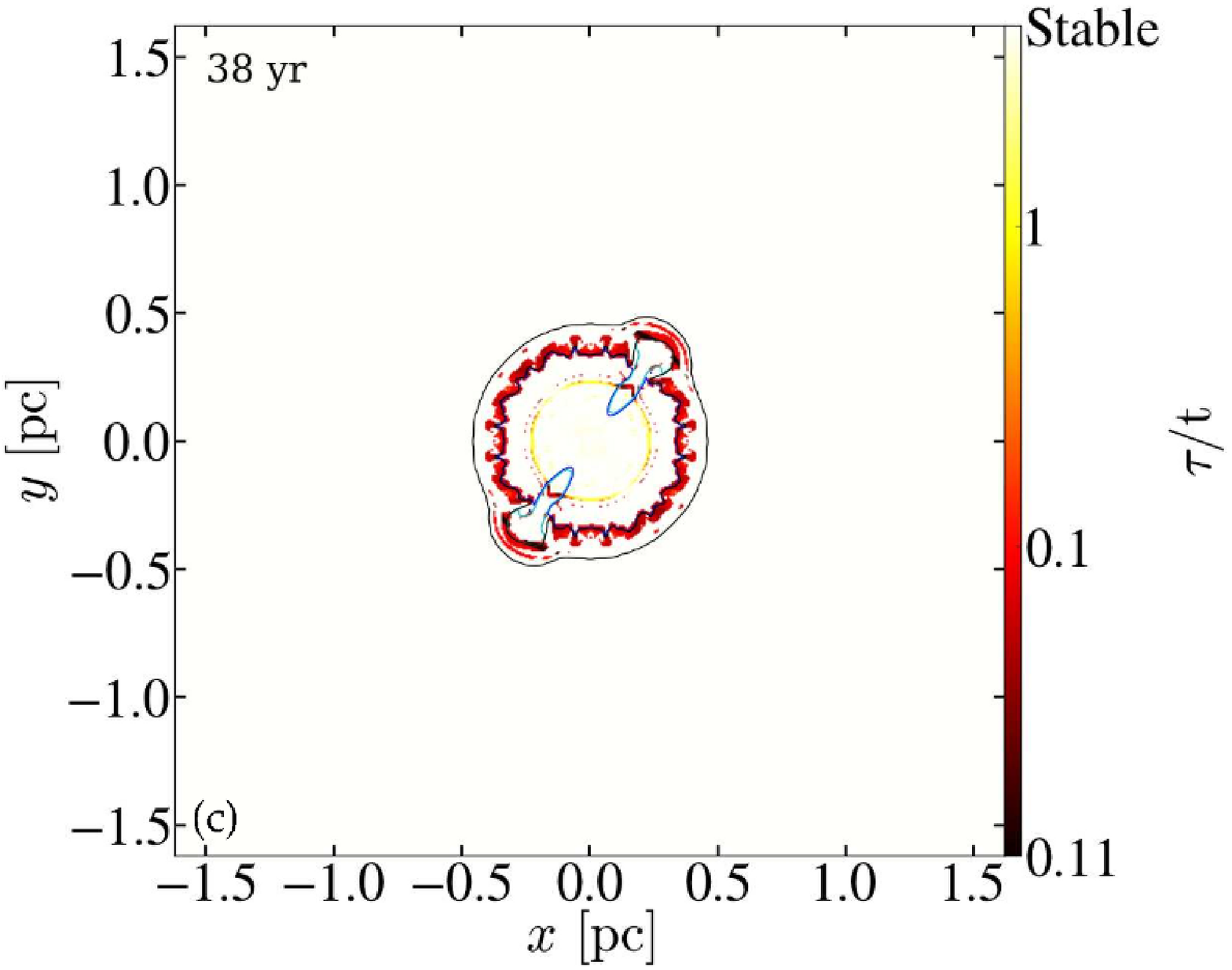}}
\hspace*{-0.5 cm}
\subfigure{\label{subfigure:jetsrt2}\includegraphics[scale=0.41,clip=false,trim=0 0 0 0]{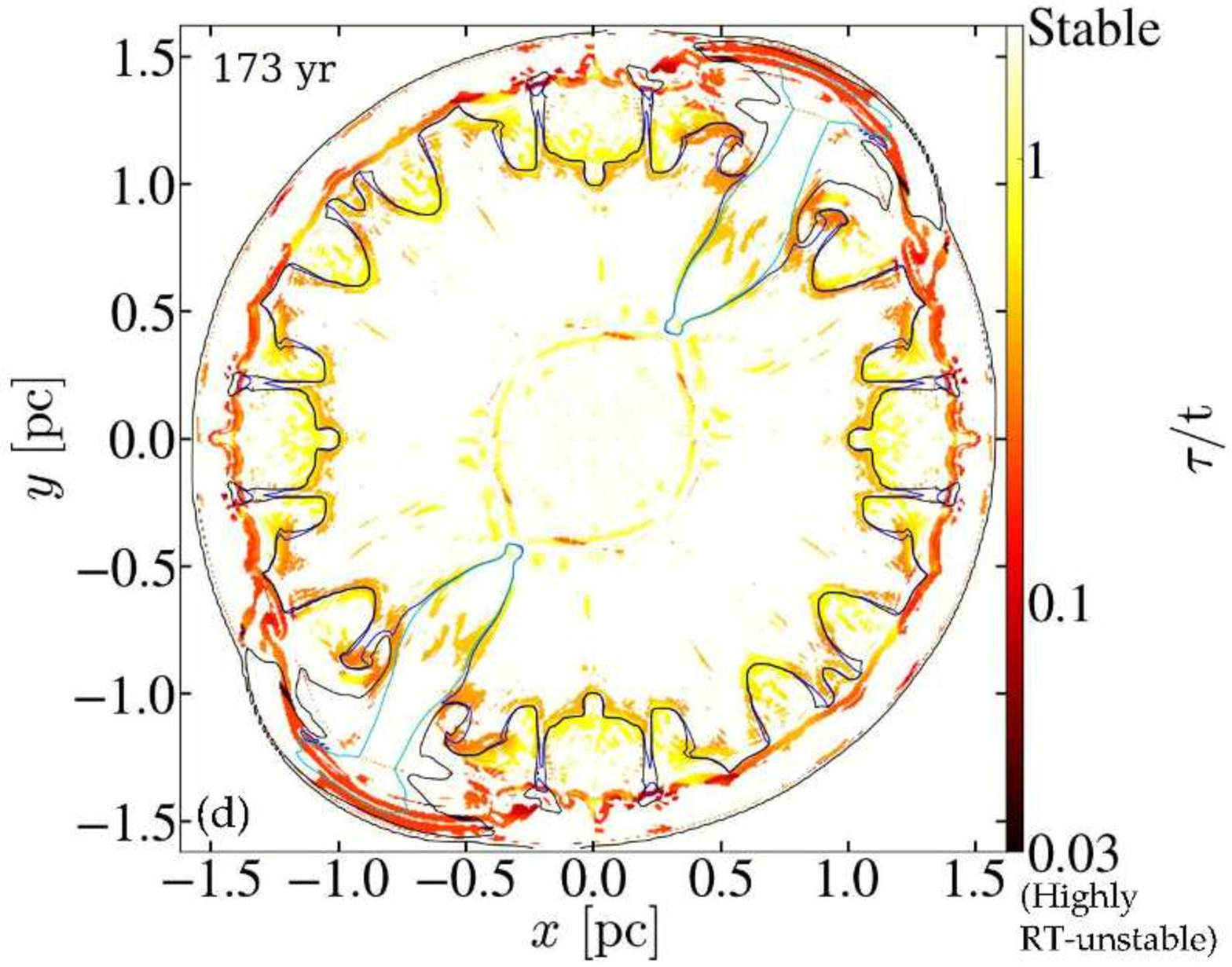}}
\caption{
Maps in the $x-y$ plane of several quantities in the spherical CSM with PEJ model.
(a) Temperature.
(b) Pressure with overlapping velocity vectors.
The vectors length is scaled linearly with the velocity, with the longest vectors corresponding to $v \simeq 10,000 ${ }\kms.
(c-d) The ratio of the Rayleigh-Taylor instability growth time to the time of the simulation
(marked on each panel) according to Eq. \ref{eq:rt}.
}
\label{fig:otherplots}
\end{center}
\end{figure}

As the SN ejecta hits the CSM, part of the kinetic energy of the ejecta is deposited in the CSM shell,
and the CSM is accelerated outward.
Two shock waves are formed.
The first shock runs outward into the CSM, and then passes to the ISM.
The second shock is a reverse shock that runs inward into the incoming ejecta.
There are two contact discontinuities in the flow,
one between the shocked ejecta and  the shocked CSM,
and the other between the CSM and ISM (initially preshock media, later shocked media).
The shocks are clearly seen in all panels of Fig. \ref{fig:densityPlots} and \ref{fig:densityPlots2},
and are marked on panel \ref{subfigure:densears2}.
By ISM we refer to the medium outside the dense CSM shell.
This ISM might contain previous stellar wind, and hence is actually a low-density CSM.
The reverse shock is best seen in the temperature and pressure maps in Fig. \ref{fig:otherplots}, and is marked on panel \ref{subfigure:jetstemperature}.

In our flow setting the shocked CSM is denser than the shocked ejecta.
Therefore, as the shocked ejecta accelerates the shocked CSM,
the flow becomes Rayleigh-Taylor (RT) unstable,
leading to the formation of `RT fingers' and the penetration of the two media into each other.
The fingers are clearly seen in Fig. \ref{fig:densityPlots}, and are marked on panel \ref{subfigure:densears2}.
Similar instabilities are present in SN Ia simulations performed by \cite{Warren2013}.
The shocked ISM is of lower density than the shocked CSM.
As the shocked ISM slows down the shocked CSM this contact discontinuity also becomes RT unstable.
These unstable regions are best presented in panels \ref{subfigure:jetsrt1} and \ref{subfigure:jetsrt2}, as described next.
The post shock CSM gas is heated up to temperatures $\rm{of} \simeq 10^9 \K$, and slowly cools down mainly via adiabatic cooling.
For our initial conditions the radiative cooling time scale of the shocked CSM is $\gtrsim 1000 \yrs$,
whereas the flow time in our simulation is up to $\simeq 170 \yrs$.

The initial departure from spherical symmetry
(see Fig. \ref{subfigure:densears1} and \ref{subfigure:densjets1})
is 'magnified' in absolute size by the sweeping SN ejecta
as the lobes are inflated from within by the ejecta-CSM interaction
(see  Fig. \ref{subfigure:densears2}, \ref{subfigure:densjets2} and \ref{fig:densityPlots2}).
In both models, a large-scale deviation from spherical symmetry is formed along the symmetry axis.
The outflow in the inner part of the SNR at late simulation times is mostly governed by the low-velocity ejecta
that was located close to the center of the SN explosion in the beginning of the simulation.
Our modeling of this slow ejecta is not accurate due to numerical limitations (pixel size),
thus the simulation results of the innermost regions (up to $0.5 \pc$ at $173 \yr$) of the SNR are of limited accuracy.
The overall "clumpiness" of the regions dominated by RT fingers resembles the fleecy structure analysed in \cite{Warren2013} for the interiors of type Ia SNRs.
As the differences between the results of the CSM-lobes (Fig. \ref{subfigure:densears1},b) model
and the PEJ model (Fig. \ref{subfigure:densjets1},d) are small, we choose to focus on the PEJ model for further analysis.

In Fig. \ref{fig:densityPlots2} we present the density maps in different planes,
and in Fig \ref{fig:otherplots} we present more physical quantities of the flow.
Both figures emphasize the main features of the flow, in particular the structure and evolution of the instabilities.
Fig. \ref{subfigure:jetstemperature} and \ref{subfigure:jetspressure} show temperature and pressure maps.
Fig. \ref{subfigure:jetsrt1} and \ref{subfigure:jetsrt2} show the ratio
of the RT-growth time $\tau_{\rm RT}$, to the present time of the simulation $t_{\rm sim}$.
This ratio is calculated as
\begin{equation}
\label{eq:rt}
\frac{\tau_{\rm RT}}{t_{\rm sim}} = \frac{1}{t_{\rm sim}} \sqrt{\frac{\lambda \rho}{\vert \vec{\nabla} P \vert}},
\end{equation}
where $\lambda$ is the typical size of the RT instabilities
and $\vec{\nabla} P$ is the pressure gradient in the $x-y$ plane.
We scale $\lambda$ as a fraction of the initial CSM shell width, $0.03 \pc$,
and take it to be, somewhat arbitrarily, $0.01 \pc$.
The exact value is of no significance for our analysis.
$\lambda$ equals 1.6 times cell size.

\begin{figure}[h!]
\begin{center}
\hspace*{-1.0 cm}
\subfigure{\label{subfigure:denslow}\includegraphics[scale=0.41,clip=false,trim=0 0 0 0]{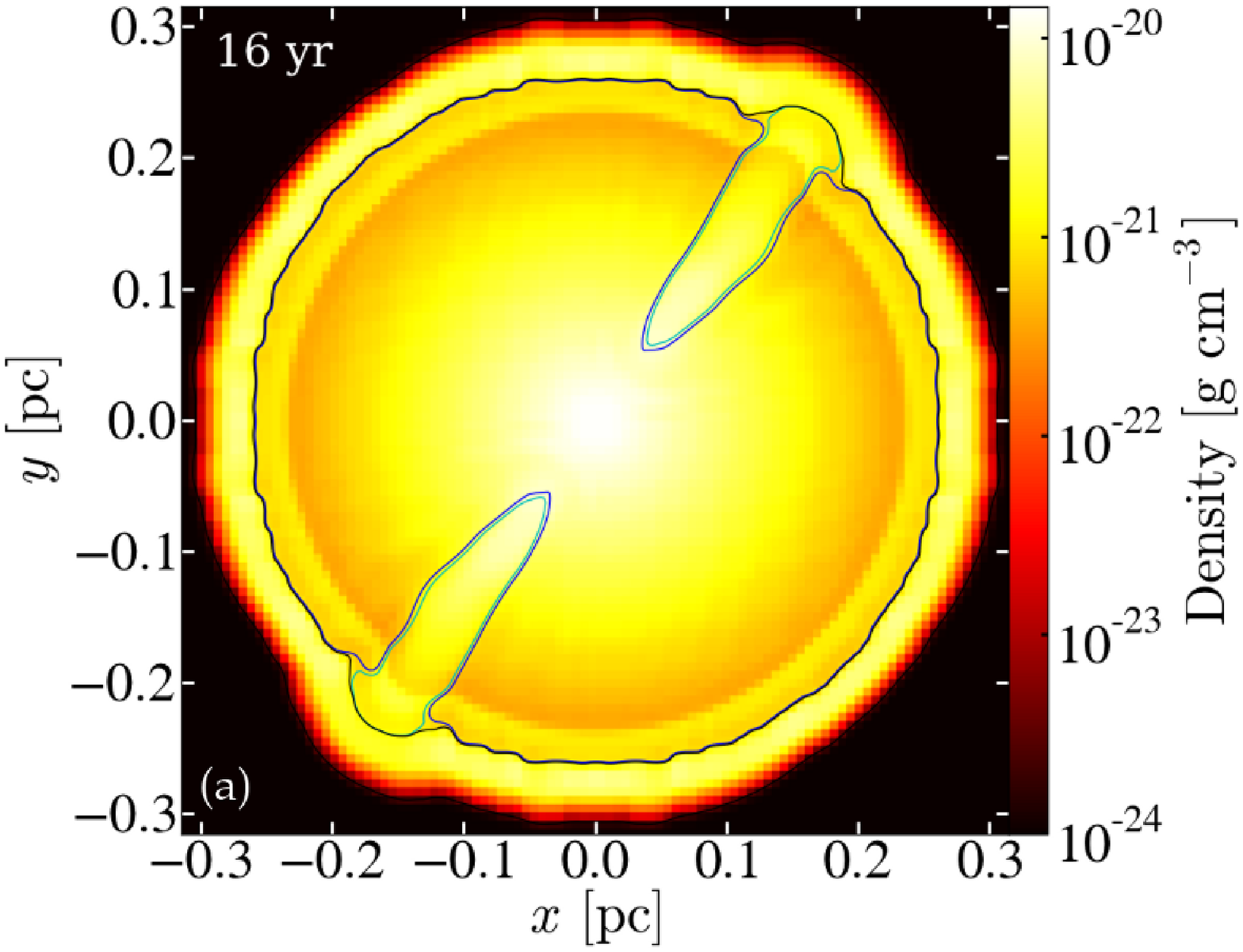}}
\subfigure{\label{subfigure:denshigh}\includegraphics[scale=0.41,clip=false,trim=0 0 0 0]{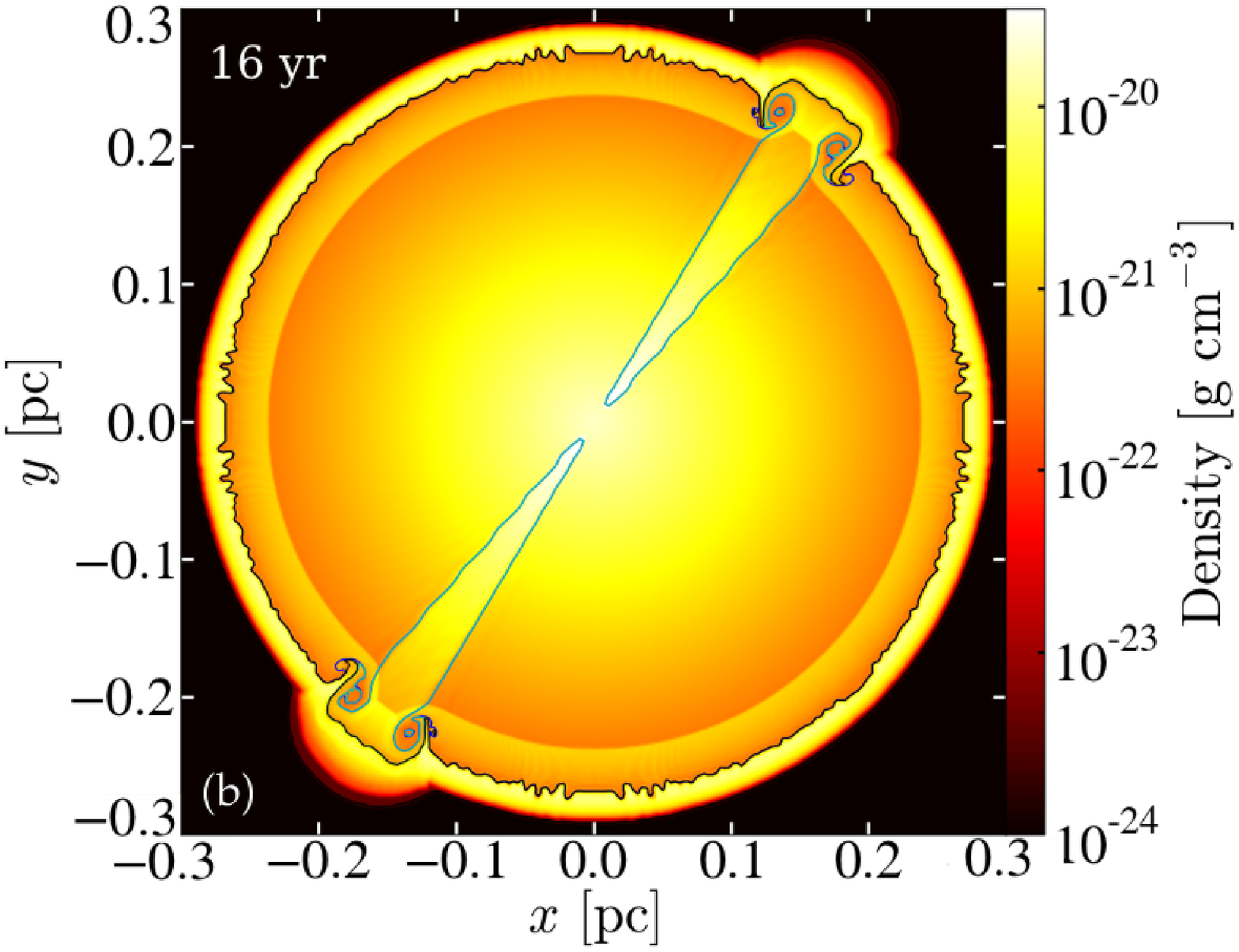}}\\
\hspace*{-0.9 cm}
\subfigure{\label{subfigure:temphigh}\includegraphics[scale=0.397,clip=false,trim=0 0 0 0]{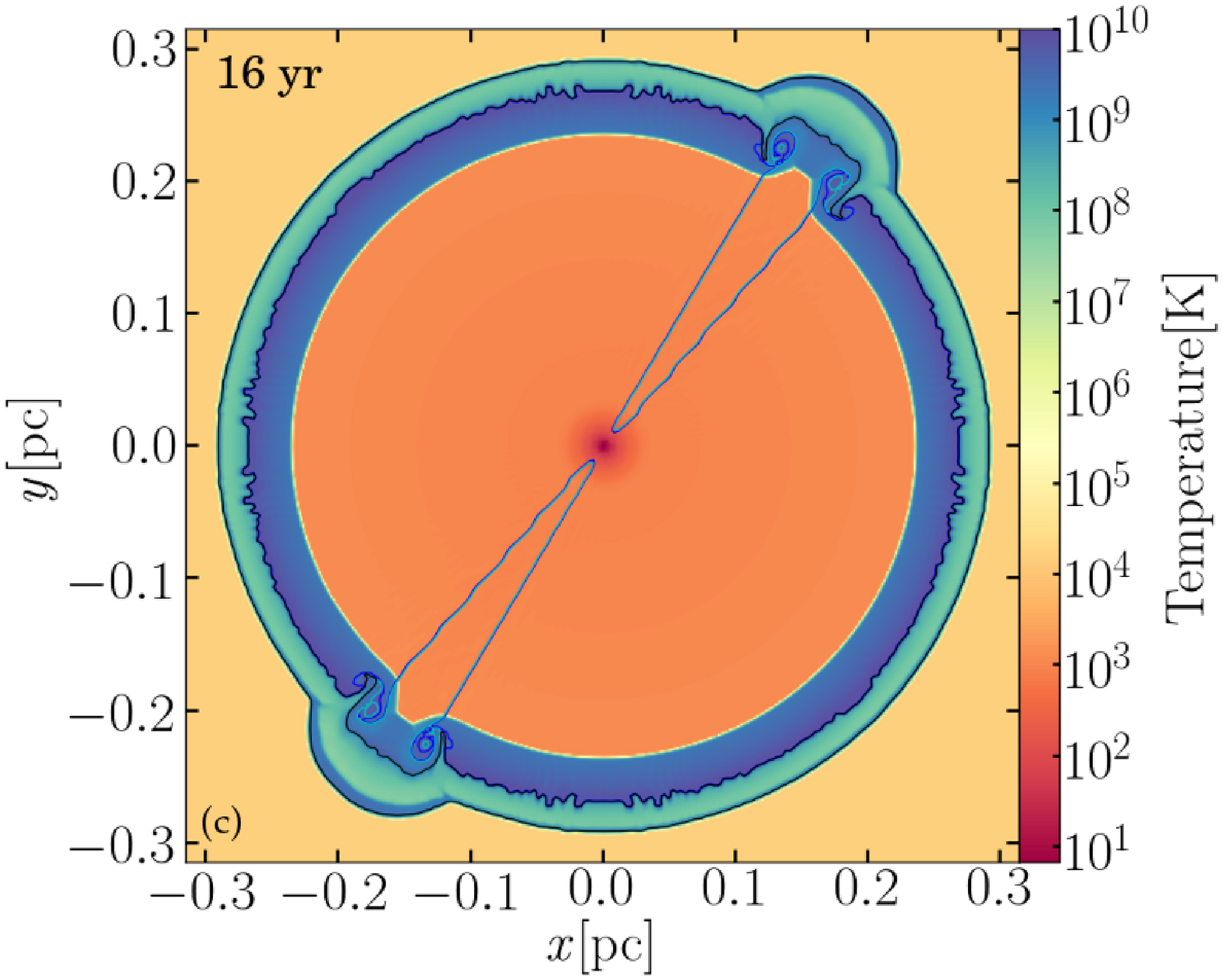}}
\hspace{0.2 cm}
\subfigure{\label{subfigure:rthigh}\includegraphics[scale=0.41,clip=false,trim=0 0 0 0]{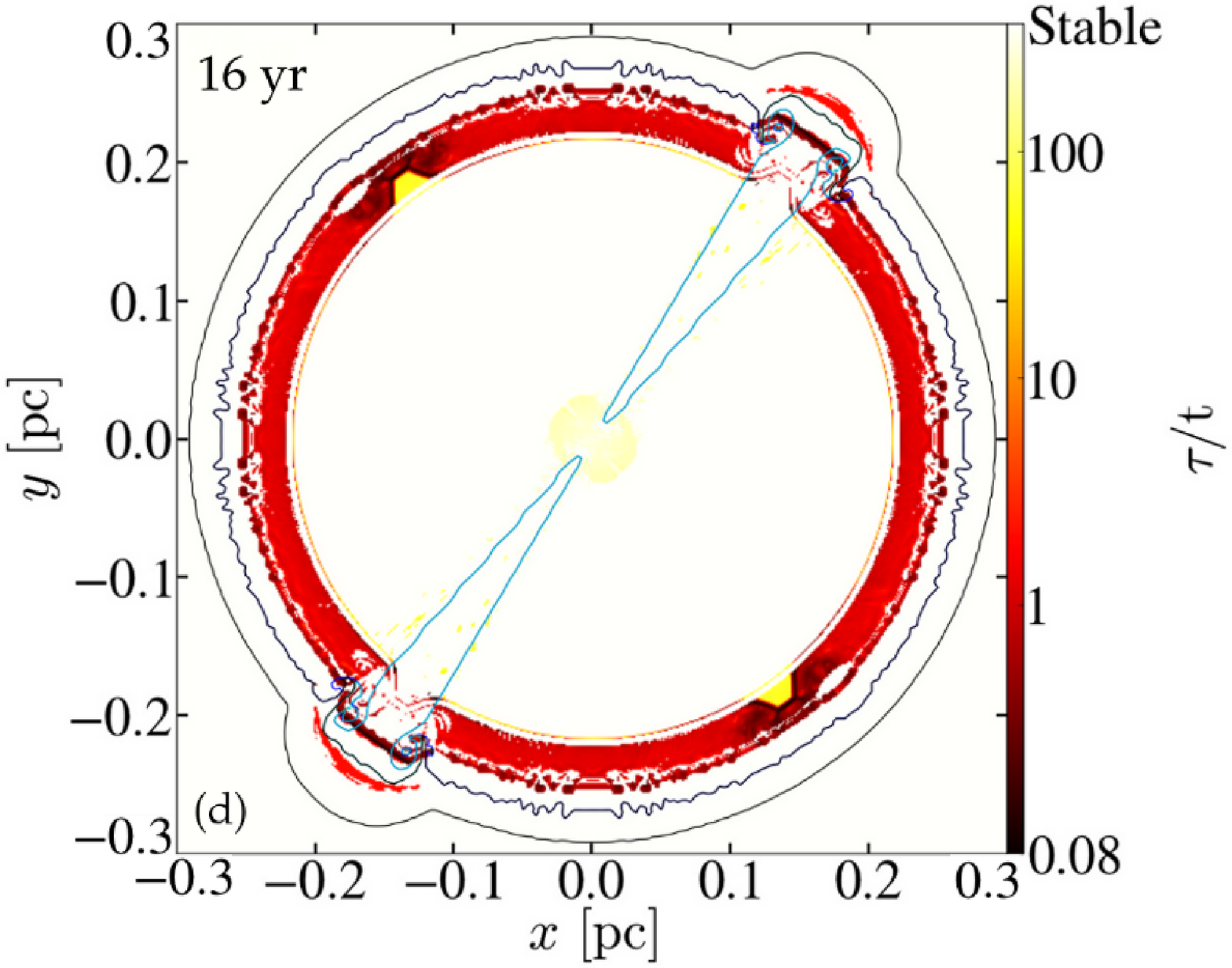}}
\caption
{Maps in the $x-y$ plane of several quantities in the spherical CSM with PEJ runs at an early time ($t = 16 \yr$).
At this time the forward shock has just passed through the CSM shell and has reached the ISM.
(a) Density map in the nominal (lower resolution) run (512 cells, $3.2 \pc$ per axis).
(b) Density map in the high-resolution run (512 cells, $0.65 \pc$ per axis).
(c) Temperature map in the high-resolution run.
(d) RT instabilities timescale in the high-resolution run.
}
\label{fig:resolutionplots}
\end{center}
\end{figure}

In addition to the nominal resolution runs whose results are presented above,
we perform a high-resolution simulation that focuses on early times
(around the time of the initial interaction of
the SN ejecta with the CSM shell), before the CSM shell expanded to higher radii.
In this run $\lambda=0.01 \pc$ equals the size of 8 numerical cells.
Fig. \ref{fig:resolutionplots} shows various quantities at early times of the high-resolution simulation,
beside panel \ref{subfigure:denslow} that presents the density map of the nominal lower resolution run presented in Fig.
\ref{subfigure:densjets1}, \ref{subfigure:densjets2}, \ref{fig:densityPlots2} and \ref{fig:otherplots}.
While there are differences between the nominal and the high-resolution runs on small
scales,  in particular the formation of RT instabilities
on the high-resolution density map (compare panel \ref{subfigure:denslow} with \ref{subfigure:denshigh}),
the two simulations are similar on large scales.
As well, we note that the 'Ears' that begin to form in the high-resolution run have 'sharper' structure than the 'Ears' in the
nominal-resolution run.

\begin{figure}[h!]
\begin{center}
\subfigure{\label{subfigure:radial1}\includegraphics[scale=0.20,clip=false,trim=0 0 0 0]{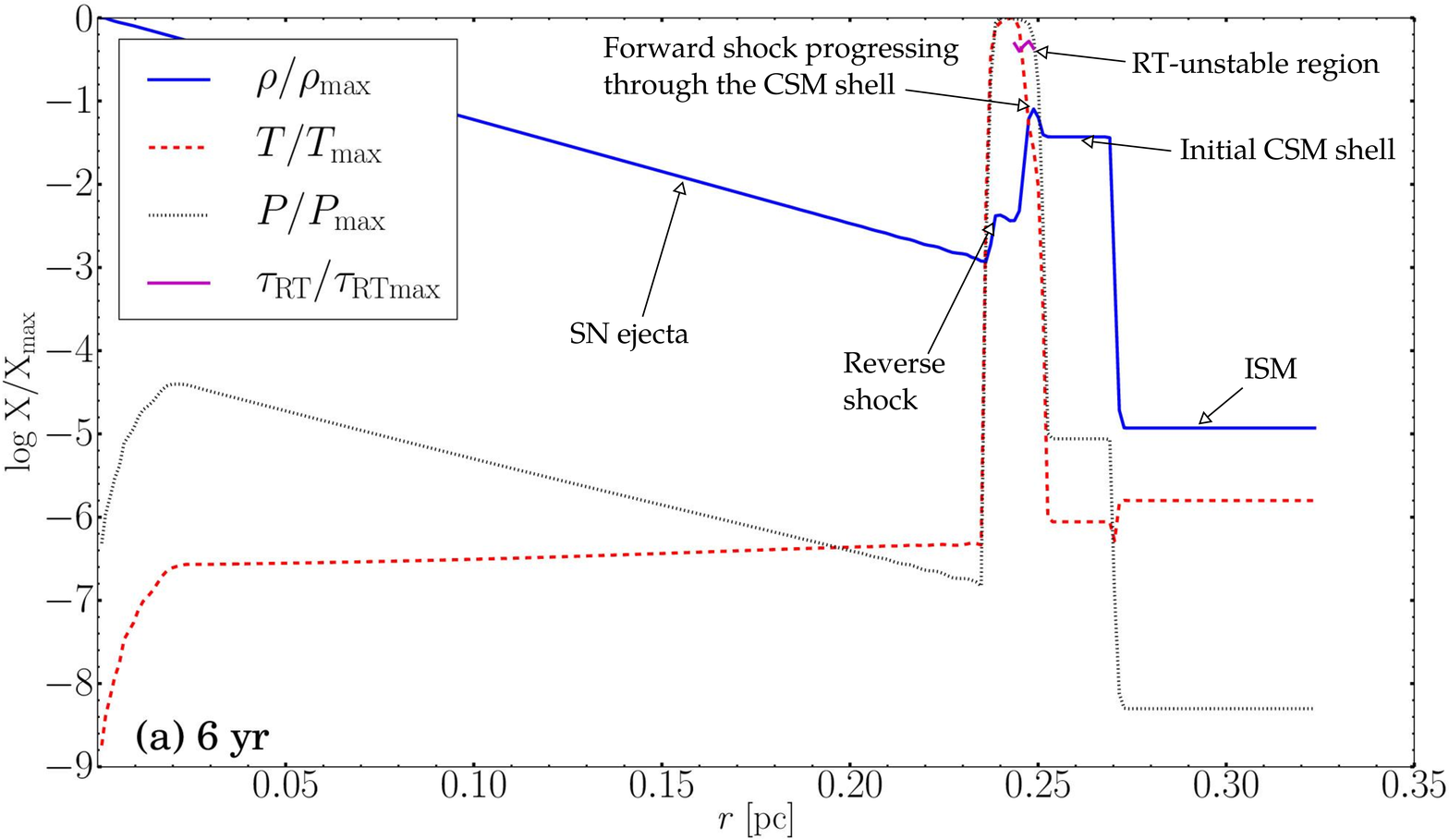}}
\subfigure{\label{subfigure:radial2}\includegraphics[scale=0.20,clip=false,trim=0 0 0 0]{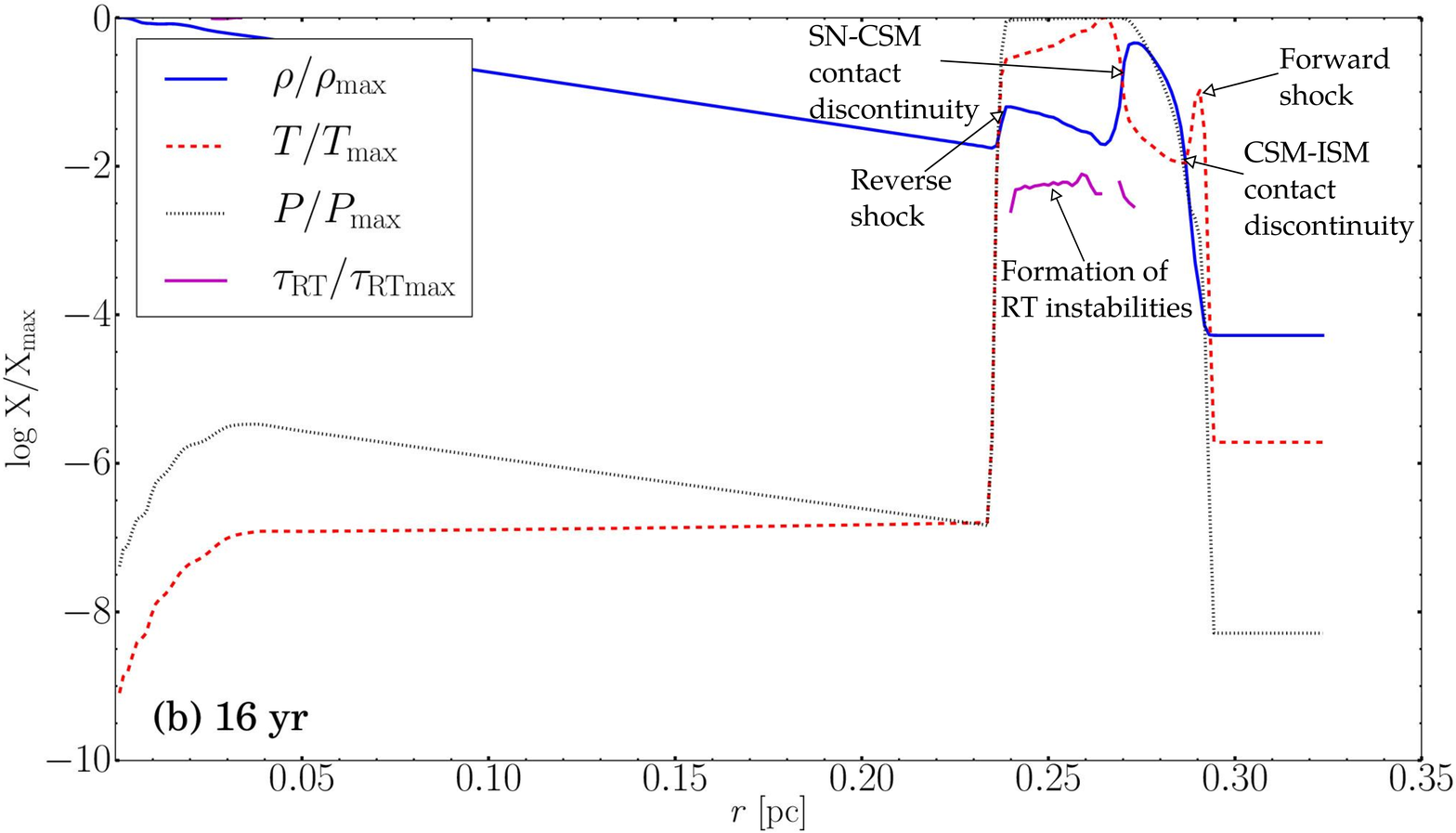}}
\caption{
Radial profiles of density (solid blue), temperature (dashed red), pressure (dotted black) and RT timescale (solid magenta),
taken from the center of the simulation grid along the positive $x$ axis at two early times.
(a) $6 \yr$. The shock is inside the CSM shell.
The maximum values are
$\rho_{\rm max} \simeq 8.5 \times 10^{-20} \g \cm^{-3}$,
$T_{\rm max} \simeq 10^{10} \K$,
$P_{\rm max} \simeq 2 \times 10^{-4} \erg \cm^{-3}$,
${\tau_{\rm RTmax}}/{t_{\rm sim}} \simeq 2$.
(b) $16 \yr$. The shock has passed the CSM shell and is in the ISM.
$\rho_{\rm max} \simeq 1.9 \times 10^{-20} \g \cm^{-3}$,
$T_{\rm max} \simeq 8.3 \times 10^{9} \K$,
$P_{\rm max} \simeq 2 \times 10^{-4} \erg \cm^{-3}$,
${\tau_{\rm RTmax}}/{t_{\rm sim}} \simeq 30$.
High-resolution PEJ run (512 cells, $0.65 \pc$ per axis).
Note that a lower value of $\tau_{\rm RT}$ means faster growth of instabilities.
}
\label{fig:radialprofiles}
\end{center}
\end{figure}

The early times, when the forward shock front is still inside the CSM,
are of particular interest for understanding the interaction between the media.
Fig. 7 shows the radial profiles of the density, temperature, pressure and the RT instability growth time, at two
early times.
In panel \ref{subfigure:radial1} we present the profiles at $t=6 \yr$,
when the forward shock front is still inside the CSM (not all the CSM has been shocked yet),
while in panel \ref{subfigure:radial2} at $t=16 \yr$, the shock front is running through the ISM.
The profiles are drawn from the center of the grid along the positive $x$ axis.
The formation of the RT instabilities begins at an early time in the simulation,
when the SN ejecta first hit the CSM shell.
At this time the CSM shell did not move much from its original location
at radii of $0.24-0.27 \pc$ (marked in panel \ref{subfigure:radial1}).
However, the shell's inner part is hit by the
SN ejecta and a shock runs into the CSM and heats the material up to $T_{\rm max} \simeq 10^{10} \K$.
As the shock progresses further into the shell,
a highly RT-unstable region develops around $0.24-0.26 \pc$ (marked in panel \ref{subfigure:radial2}).
This region is on the contact discontinuity between the shocked ejecta
and the shocked CSM ($0.28 \pc$, where a density jump is seen).
Over time, these instabilities develop to the large fingers seen in Fig. \ref {subfigure:densears2}, \ref{subfigure:densjets2}
and \ref{fig:densityPlots2}.

The main results of our simulations presented above are as follows.
(1) In both the CSM-lobes model and the PEJ model two 'Ears' on opposite sides of the SNR are formed.
These protrusions are of significant size
($r \simeq 0.5 \pc$ at $170 \yrs$ for the parameters we used), and may appear as ‘Ears’ in observations.
In the next section we examine whether the assumed jets can account for the structure of some SNRs.
(2) The interaction of the SN ejecta with the CSM shell creates a complex density structure inside the SNR,
mainly through evolution of Rayleigh-Taylor instabilities.
\section{COMPARISON WITH SNRs}
\label{sec:comparison}
We attempt to match some of the features in our numerical results with some morphological features
in the SNRs Kepler and G299.2-2.9.
The X-ray emission from the asymmetrical Kepler's SNR has been
studied in the past, recently by \cite{Burkey2013}.
They interpret the observed distribution of CSM in Kepler's SNR
as resulting from a disk seen edge-on, which extends from the center of the SNR
along the line that marks our proposed symmetry axis in Fig. \ref{fig:observations}.
Our interpretation of the observations is different,
in that we propose that the line connecting the two 'Ears' is actually a symmetry axis and not the equatorial plane.
Our conjecture of the direction of the symmetry axis is based on similar interpretations
applied to planetary nebulae that possess similar 'double-Ear' morphologies (see Section \ref{sec:numerical}).
Following these interpretations we here attribute the 'Ears' observed in SNRs to jets.
In particular, we examine the Kepler and G299.2-2.9 SNRs.

\begin{figure}[h!]
\begin{center}
\includegraphics*[scale=0.5,clip=false,trim=0 0 0 0]{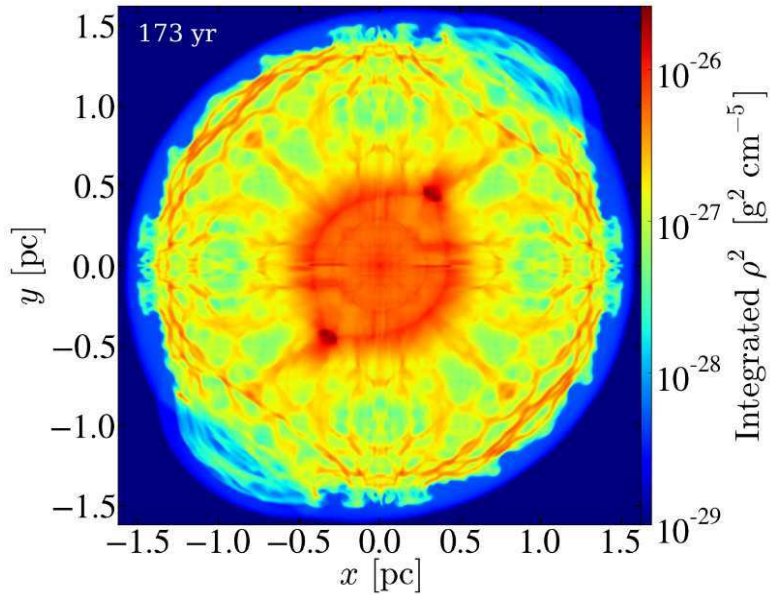}
\caption{
Integrated $\rho^2$ for the spherical CSM with PEJ run.
A density map of the same run is presented in panel \ref{subfigure:densjets2}.
Apparent are the two 'ears' features, formed by jets close to the time of the SN explosion.}
\label{fig:integratedDens}
\end{center}
\end{figure}

The exact reconstruction of the expected x-ray emission map from our simulations is beyond the scope of this paper.
However, as x-ray emission magnitude is proportional to the square of the density
we can make a qualitative comparison by constructing an `intensity map'
by simply integrating $\rho^2$ along the line of sight
(a similar approach was used in \citealt{Burkey2013}).
We take the 'Ears' to lie in the plane of the sky.
This intensity map is shown in Fig. \ref{fig:integratedDens}.

Considering our results presented in Section \ref{sec:results} for both the CSM-lobes and the PEJ models
at a late simulation time (e.g. the resulting density maps shown in panels \ref{subfigure:densears2} and \ref{subfigure:densjets2})
and making a qualitative comparison between
Fig. \ref{fig:integratedDens} and the observed images in Fig. \ref{fig:observations},
we are led to suggest that both our proposed models - the CSM-lobes model and the PEJ model -
might explain the observed 'Ears' features in Kepler SNR and G299.2-2.9 SNR.
\section{SUMMARY}
\label{sec:summary}

Although many SN Ia remnants exhibit almost spherical large-scale symmetry,
there are several remnants of SN Ia (SNRs) that show an axisymmetrical deviation from spherical large-scale structure.
The Kepler and G299.2-2.9 SNRs, for example, have two protrusions ('Ears') on the outer rim of the SNR,
positioned exactly opposite to each other (see Fig. \ref{fig:observations}).
Noting that some planetary nebulae (PNs) have similar axisymmetrical morphologies with two opposite 'ears',
and following interpretations that attribute these features to jets,
we propose that the 'Ears' in these two SNRs can also be attributed to jets.

In Section \ref{sec:numerical} we discussed two scenarios for the formation of such protrusions:
(a) The CSM-lobes model, in which a SN explosion takes place inside an overall spherical shell with two hollow small lobes,
mimicking the structure of such two opposite small lobes observed in some elliptical PNs.
This model is plausible if the SN explosion takes place via the SD or the CD scenarios.
(b) The pre-explosion jets (PEJ) model, in which the SN explosion is ignited inside a
CSM shell (which might be spherical or not),
with two jets added inside the SN ejecta.
This can occur only in the CD scenario during the core-WD merger process \citep{Soker2013}.
The models are described schematically in Fig. \ref{fig:schemes}.

The initial conditions of the runs were motivated by the Kepler SNR,
although we did not try to reproduce its 'Ears' one to one.
We employed 3D hydrodynamical simulations to follow the interaction of the SN ejecta
with the CSM in the two scenarios - CSM-lobes and PEJ.

The main results of our simulations are presented in Section \ref{sec:results}, and can be summarized as follows.
(1) In both the CSM-lobes model and the PEJ model two 'Ears' on opposite sides of the SNR are formed.
These protrusions are of significant size
($r \simeq 0.5 \pc$ at $170 \yrs$ for the parameters we used), and may appear as `Ears' in observations.
(2) The interaction of the SN ejecta with the CSM shell creates a complex density
structure inside the SNR, mainly through evolution of Rayleigh-Taylor instabilities.

In Section \ref{sec:comparison} we compare the results of our simulations with observations,
and propose that both our suggested scenarios (CSM-lobes and PEJ)
may explain the observed morphologies of the Kepler and G299.2-2.9 SNRs.
In general, we propose that some SN Ia may blow jets close to the time of the explosion,
offering an explanation to the occasionally observed axisymmetrical morphology for the remnants of such SN.

\vspace{1 cm}
We thank Alexei Baskin for useful comments. 
This research was supported by the Asher Fund for Space Research at the Technion, the E. and J. Bishop Research Fund
at the Technion, and the USA-Israel Binational Science Foundation.
The FLASH code used in this work is developed in part by the US Department of Energy under Grant No.
B523820 to the Center for Astrophysical Thermonuclear Flashes at the University of Chicago.
The simulations were performed on the TAMNUN HPC cluster at the Technion.

\end{document}